\documentclass[10pt, conference, compsocconf]{IEEEtran}
\usepackage{color,graphicx}
\usepackage{circuitikz}
\usepackage{amsmath}
\usepackage{amssymb}
\usepackage{amsthm}
\usepackage{subsections}
\usepackage[ruled]{algorithm2e}
\usepackage{enumerate}
\usepackage{syntonly}
\usepackage{hhline}
\usepackage{multicol}
\usepackage{array}
\usepackage[tight,footnotesize]{subfigure}
\usepackage{stfloats}
\usepackage{url}
\usepackage{threeparttable}
\usepackage{cite}
\usepackage{multirow}
\usepackage{longtable}
\usepackage{url}
\usepackage[breaklinks]{hyperref}
\usepackage[hyphenbreaks]{breakurl}

\usepackage{todonotes}
\usepackage{tikz}
\usepackage{rotating}
\usepackage{amsmath,amssymb,amsthm}
\usepackage{listings}
\usepackage{todonotes}
\usepackage{ wasysym }
\usepackage{pifont}
\usepackage{csvsimple}
\usepackage{siunitx}
\usepackage{tabularx}
\usepackage{xifthen}
\usepackage[T1]{fontenc}
\usepackage{stmaryrd}
\IEEEoverridecommandlockouts

\usetikzlibrary{arrows}
\usetikzlibrary{positioning}
\usetikzlibrary{shadows}
\usetikzlibrary{arrows,backgrounds,calc,chains,
	matrix,positioning,shapes,shapes.geometric,
	shapes.arrows, decorations.pathmorphing, decorations.pathreplacing}

\tikzstyle{vecArrow} = [thick, decoration={markings,mark=at position
	1 with {\arrow[semithick]{open triangle 60}}},
double distance=1.4pt, shorten >= 5.5pt,
preaction = {decorate},
postaction = {draw,line width=1.4pt, white,shorten >= 4.5pt}]

\newtheorem{example}{Example}

\usepackage{etoolbox}
\makeatletter
\patchcmd{\maketitle}{\@copyrightspace}{}{}{}
\makeatother

\usetikzlibrary{backgrounds,calc,chains,matrix,positioning,shapes,shapes.geometric,shapes.arrows}

\tikzset{%
	font={\footnotesize},
	vertex/.style={draw,circle,inner sep=0pt,minimum width=0.5cm,minimum height=0.5cm},
	zeroterm/.style={below,inner sep=0pt,font=\tiny}
}

\author{}

\author{     
   Samah Mohamed Saeed, 
   Xiaotong Cui,
   Robert Wille,
   Alwin Zulehner,
   Kaijie Wu,
   Rolf Drechsler,
 \\and Ramesh Karri, Senior Member, IEEE 

\thanks{Samah Mohamed Saeed is with the Institute of Technology, University of Washington, Tacoma, WA, USA (email: samahs@uw.edu).}
\thanks{Xiaotong Cui is with the College of Computer Science, Chongqing University, Chongqing 400044 China.}
\thanks{Robert Wille and Alwin Zulehner are with the Institute for Integrated Circuits, Johannes Kepler University Linz, 4040 Linz, Austria.}
\thanks{Ramesh Karri and Kaijie Wu are with Tandon School of Engineering, New York University, NY, USA.}
\thanks{Rolf Drechsler is with the University of Bremen, 28359 Bremen, Germany and also with DFKI GmbH, 28359 Bremen, Germany.}

}

\begin{document}

\title{Towards Reverse Engineering Reversible Logic}
\maketitle

\subsection*{Abstract}

Reversible logic has two main properties. First, the number of inputs is equal to the number of outputs. Second, it implements a one-to-one mapping; i.e., one can reconstruct the inputs from the outputs. These properties enable its applications in building quantum computing architectures. 

In this paper we study reverse engineering of reversible logic circuits, including reverse engineering of non-reversible functions embedded into reversible circuits. 
We propose the number of embeddings of non-reversible functions into a reversible circuit as the security metric for reverse engineering. We analyze the security benefits of automatic synthesis of reversible circuits. We use our proposed security metric to show that the functional synthesis approaches yield reversible circuits that are more resilient to reverse engineering than the structural synthesis approaches. Finally, we propose scrambling of the inputs and outputs of a reversible circuit to thwart reverse engineering. 
\maketitle

\begin{IEEEkeywords}
Reversible logic, Reverse engineering, Security, Structural synthesis, Functional synthesis, BDD synthesis, QMDD synthesis, Number of embeddings.

\end{IEEEkeywords}

\section{Introduction}
\label{sec:introduction}

The globalization of the design flow for integrated circuits~(ICs) leads to security vulnerabilities such as reverse engineering \cite{Torrance:2009:SIR:1617722.1617758}, IC counterfeiting, Intellectual Property~(IP) piracy, unauthorized overproduction
by the contract foundry~\cite{1628506,4484823}, and  malicious circuit modification~\cite{tehranipoor2010survey}. An adversary anywhere in the design flow can reverse engineer the IP/IC, steals its ownership, and make pirated copies. An untrusted foundry can overbuild ICs and distribute them illegally. Furthermore, detailed knowledge of the design can allow one to identify sensitive parts of the design and make malicious modifications, refereed to as Hardware Trojans.  

Reverse engineering can discover the technology used in the device~\cite{Rev_tech1}, extract the gate level implementation~\cite{Torrance:2009:SIR:1617722.1617758} and reveal the functionality of the design~\cite{Rev_ref_JV}. Software tools are used to reverse engineer the chip~\cite{Rev_tool1_JV,Rev_tool2_JV}. An adversary can use reverse engineering to steal IP, illegally fabricate ICs and insert Hardware Trojans. Design-for-Trust techniques thwart reverse engineering of CMOS-based logic circuits. Obfuscation techniques have been proposed to hide the implementation and the functionality of the design by inserting additional gates, which are controlled by a secret key~\cite{5604160,6616532,7362173}. The design can function correctly only when the correct key is applied to these gates. IC camouflaging is a layout-level protection against malicious end users from extracting the gate level implementation of the design~\cite{chow2012camouflaging,Rajendran:2013:SAI:2508859.2516656, 6651879}. In camouflaging, layouts of the standard logic gates are designed to look alike. Most of these considerations focus on conventional CMOS-based technologies.

\emph{Reversible logic} is a novel computing paradigm where one obtains not only the output value for a given input value but also the other way around. Obviously, reversible logic significantly differs from conventional CMOS-based non-reversible logic. In traditional CMOS-based logic, it is possible to infer some of the inputs of a conventional NAND gate if its output is 0 (then, both inputs are 1) while it is not possible to unambiguously infer the input values if the NAND gate output is a 1. Reversible circuits realize bijective  $n$-input $n$-output functions that map each possible input vector to a unique output vector. This is beneficial in building quantum computing architectures~\cite{2000:QCQ:544199} since quantum computations are inherently reversible~\cite{BBC+:95,DBLP:conf/rc/NiemannBCJW15}. 

Besides, reversible logic has applications in low power computing~\cite{Landauer61,Ben:73},  adiabatic computing~\cite{363692}, encoder/decoder design ~\cite{ZulehnerW17,DBLP:conf/date/WilleDOO12,DBLP:conf/date/WilleKHWO16}, circuit verification~\cite{DBLP:conf/date/AmaruGW16}, and optical computing~\cite{Cuykendall:87}.

Security assessment of reversible circuits is the focus of this paper. Reversible circuits differ significantly from conventional circuits and hence may be susceptible to these and other threats. For example, in reversible circuits fanout and feedback are not (directly) allowed and each circuit is realized as a cascade of reversible gates. Similarly, the function (which is not necessarily reversible) is embedded into a reversible circuit structure resulting in ancillary inputs and garbage outputs. We investigate how these unique constraints inform the security of reversible circuits with a particular focus on reverse engineering. We investigate the challenges for a reverse engineer assuming different levels of knowledge about the synthesis schemes. Next, we propose a metric to quantify the ability of a design to resist reverse engineering. Finally, we propose scrambling the inputs and outputs to make reverse engineering difficult. Overall, this paper provides a first understanding of the risks of reverse engineering in reversible logic.

The remainder of this paper is organized as follows. Section~II provides a background on reversible logic. Then, reverse engineering of reversible circuits is analyzed in detail for different threat models in Section III. In Section IV, we provide a simple input/output scrambling solution to thwart reverse engineering of reversible circuits. Experimental results in Section~V demonstrate the difficulty of reverse engineering of non-reversible functions embedded into reversible circuits implementations and report the number of possible embeddings encountered by an attacker. We evaluate the proposed scrambling technique with respect to the number of possible embeddings and the costs. We conclude the paper in Section~VI.

\section{Background}
In this section, we provide an overview of reversible logic and established automatic synthesis approaches to realize a (not necessarily reversible)
target function using reversible logic gates. 
\subsection{Reversible Logic}
A reversible logic circuit has an equal number of input and output signals. Furthermore, the reversible circuit realizes a bijection, i.e.~each input assignment maps to a unique output assignment. Accordingly, computations can be performed in both directions (from the inputs to the outputs and vice versa).

Reversible circuits are implemented as cascades of reversible gates. Each reversible gate over the inputs \mbox{$X=\{x_1,\dots , x_n\}$} consists of a (possibly empty) set $C_i \subseteq \{x_j \mid x_j \in X \} \cup \{\overline{x}_j \mid x_j \in X \}$ of positive ($x_j$) and negative  ($\overline{x}_j$) \emph{control lines} and a set $T\subset X\setminus C$ of \emph{target lines}. The most commonly used reversible gate is the \emph{Toffoli} gate $TOF(C,x_t)$~\cite{Tof:1980}, which consists of a single target line~$x_t\in X\setminus C$  whose value is inverted if all values on the positive (negative) control lines are set to~$1$ ($0$) or if \mbox{$C=\emptyset$}. All remaining values are passed through the gate unaltered. The cost of a reversible circuit is defined either by the number of gates or by so called \emph{quantum cost}~\cite{PhysRevA.52.3457}. The quantum cost of a Toffoli gate with $C$ positive control lines is computed as $2^{C+1}-3$, while for negative control lines, the quantum cost is computed in the same way except for the case where the Toffoli gate is entirely composed of negative control lines in which the cost is increased by two~\cite{4378213}.

\begin{figure}[htbp]
	\centering
	\begin{tikzpicture}[scale=1, every node/.style={scale=0.9}]
	\draw[line width=0.300000] (0.250000,1.250000) -- (2.75000,1.250000);
	\draw (0.1500000,1.250000) node [left] {$x_3$};
	\draw (2.85000,1.250000) node [right] {$x_3'$};
	\draw[line width=0.300000] (0.250000,0.750000) -- (2.75000,0.750000);
	\draw (0.150000,0.750000) node [left] {$x_2$};
	\draw (2.85000,0.750000) node [right] {$x_2'$};
	\draw[line width=0.300000] (0.250000,0.250000) -- (2.75000,0.250000);
	\draw (0.150000,0.250000) node [left] {$x_1$};
	\draw (2.850000,0.250000) node [right] {$x_1'$};
	
	\draw (0.375,1.250000) node [above, anchor=south] {0};
	\draw (0.375,0.750000) node [above, anchor=south] {0};
	\draw (0.375,0.250000) node [above, anchor=south] {1};

	\draw[line width=0.300000] (0.750000,0.250000) -- (0.750000,1.450000);
	\draw[line width=0.3] (0.75,1.25) circle (0.2);
	\draw[fill] (0.750000,0.250000) circle (0.100000);
	\draw (0.750000,0000) node [below] {$g_1$};
	
	\draw (1.125,1.250000) node [above, anchor=south] {1};
	\draw (1.125,0.750000) node [above, anchor=south] {0};
	\draw (1.125,0.250000) node [above, anchor=south] {1};
	
	\draw[line width=0.300000] (1.50000,0.050000) -- (1.50000,1.250000);
	\draw[fill] (1.50000,1.250000) circle (0.100000);
	\draw[line width=0.3] (1.50000,0.250000) circle (0.2);
	\draw[fill=white] (1.50000,0.750000) circle (0.100000);
	\draw (1.5000,0) node [below] {$g_2$};
	
	\draw (1.875,1.250000) node [above, anchor=south] {1};
	\draw (1.875,0.750000) node [above, anchor=south] {0};
	\draw (1.875,0.250000) node [above, anchor=south] {0};
	
	\draw[line width=0.300000] (2.250000,0.250000) -- (2.250000,.950000);
	\draw[fill] (2.250000,0.250000) circle (0.100000);
	\draw[line width=0.3] (2.250000,0.750000) circle (0.2);
	\draw (2.250000,0) node [below] {$g_3$};
	
	\draw (2.625,1.250000) node [above, anchor=south] {1};
	\draw (2.625,0.750000) node [above, anchor=south] {0};
	\draw (2.625,0.250000) node [above, anchor=south] {0};		
	
	\end{tikzpicture}
	\caption{A reversible circuit with three Toffoli gates.}
	\label{fig:rev}

\end{figure}
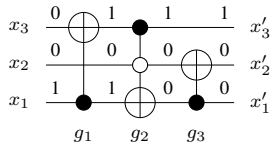	

\begin{example}
Fig.~\ref{fig:rev} shows a reversible circuit composed of three circuit lines and three Toffoli gates. The circuit is labeled with the values on the circuit lines for input $x_3x_2x_1 = 001$. The first gate $g_1 = TOF(\{x_1\}, x_3)$ inverts the value of the target line~$x_3$ since the positive control line~$x_1$ is initialized to~1. 
The second gate \mbox{$g_2 = TOF(\{x_3, \overline{x}_2\}, x_1)$} inverts the value of the target line~$x_1$. In contrast, the third gate \mbox{$g_3 = TOF(\{x_1\},x_2 )$} keeps the value of the target line~$x_2$ intact, since the positive control line~$x_1$ is 0. 
\end{example}

In order to realize non-reversible functions, \emph{ancillary inputs} and \emph{garbage outputs} are used. An ancillary input of a reversible circuit is an input that is set to a fixed value (either 0 or 1). A garbage output of a reversible circuit is an output, which is a don't care for all possible input conditions.

\begin{figure}[t]
\centering
\includegraphics[width=2.0in]{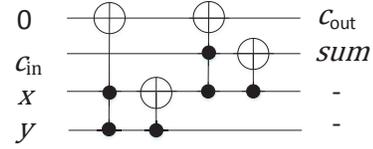}
\caption{A full adder is a non-reversible function. It can be implemented using reversible Toffoli gates, one ancillary input, and two garbage outputs. }
\label{fig:ex_adder}
\end{figure}

\begin{example}
Fig.~\ref{fig:ex_adder} shows the realization of a full adder using reversible gates with the top input of the reversible circuit set to 0. The bottom two outputs are garbage outputs of this circuit.
\end{example}

\subsection{Reversible Logic Synthesis}\label{sec:syn_rev}

Several approaches to automatically synthesize non-reversible and reversible functions using reversible gates have been proposed (see e.g.~\cite{FTR:2007,WDb:2009,GWDD:2009, SWH+:2012, DBLP:journals/jsc/SoekenTDD16, DBLP:conf/rc/SoekenDM16}). 
These synthesis approaches either implicitly or explicitly \emph{embed} the target (reversible or non-reversible)
function into a reversible function.

{\setlength{\tabcolsep}{2.5pt}
	\begin{table}[htbp]
		\centering
		\caption{Full Adder and a possible embedding into a reversible function.}
		\subfigure[Full Adder \label{wille_table:adder}]{\resizebox{0.2\textwidth}{0.1\textwidth}{\begin{tabular}{ccc|cccc}
				
				{ c$_{in}$} &    x &  y & { c$_{out}$} &  { sum} &~~ &\\
				\hline
				0 &          0 &          0 &          0 &          0 & & 0\\
				
				0 &          0 &          1 & \textbf{0} &  \textbf{1} & & \textbf{0}\\
				
				0 &          1 &          0 & \textbf{0} &  \textbf{1} & & \textbf{1}\\
				
				0 &          1 &          1 & \textit{1} & \textit{0}  & & \textit{0}\\
				
				1 &          0 &          0 & \textbf{0} & \textbf{1} & & \textbf{?}\\
				
				1 &          0 &          1 & \textit{1} & \textit{0} & & \textit{1}\\
				
				1 &          1 &          0 & \textit{1} & \textit{0} & & \textit{?}\\
				
				1 &          1 &          1 &          1 &          1 && 1\\
				
			\end{tabular}}}~~~~~~~~~\subfigure[Reversible Embedding\label{wille_table:emb_adder}]{\resizebox{0.2\textwidth}{0.1\textwidth}{\begin{tabular}{cccc|cccc}
			
			$0$      & \textbf{c$_{in}$} &    \textbf{x}    &  \textbf{y}     & \textbf{c$_{out}$} &  \textbf{sum}   &   \textbf{$g_1$}    &   \textbf{ $g_2$} \\
			\hline
			0 &          \textbf{0} &          \textbf{0} &          \textbf{0} &          \textbf{0} &          \textbf{0} &          0 &          0 \\
			
			0 &          \textbf{0} &          \textbf{0} &          \textbf{1} &          \textbf{0} &          \textbf{1} &          1 &          1 \\
			
			0 &          \textbf{0} &          \textbf{1} &          \textbf{0} &          \textbf{0} &          \textbf{1} &          1 &          0 \\
			
			0 &          \textbf{0} &          \textbf{1} &          \textbf{1} &          \textbf{1} &          \textbf{0} &          0 &          1 \\
			
			0 &          \textbf{1} &          \textbf{0} &          \textbf{0} &          \textbf{0} &          \textbf{1} &          0 &          0 \\
			
			0 &          \textbf{1} &          \textbf{0} &          \textbf{1} &          \textbf{1} &          \textbf{0} &          1 &          1 \\
			
			0 &          \textbf{1} &          \textbf{1} &          \textbf{0} &          \textbf{1} &          \textbf{0} &          1 &          0 \\
			
			0 &          \textbf{1} &          \textbf{1} &          \textbf{1} &          \textbf{1} &          \textbf{1} &          0 &          1 \\
			
			1 &          0 &          0 &          0 &          1 &          0 &          0 &          0 \\

		\end{tabular}}}
	\end{table}
}

\begin{example}
 Consider the full adder in Table~\ref{wille_table:adder}. The full adder has the carry in $c_{in}$ and summands $x$ and~$y$ as the inputs and the carry out $c_{out}$ and the \emph{sum} as the outputs. The full adder is not reversible since (1)~the number of inputs differs from the number of outputs and (2)~there is no unique input-output mapping. Adding an additional output to the function does not make it reversible. The first four rows of the truth table can be embedded with respect to reversibility as shown in the rightmost column of Table~\ref{wille_table:adder}. However, since 
$c_{out}=0$ and $sum=1$ already appeared twice (marked in bold), a unique embedding for the fifth row of the truth table is no longer possible. The same holds for the italicized rows.
\end{example}

Overall, at least $\lceil log(m)\rceil$ garbage outputs are required to make a non-reversible function reversible, where $m$ is the maximum number of times an output pattern is repeated in the truth table of the \emph{target function}~\cite{MD:2004,DBLP:journals/corr/SoekenWKMD14,ZulehnerW17Emb}. 
Since three output patterns are repeated in the full adder truth table, $\lceil log(3)\rceil =2$ garbage outputs and, hence, one additional ancillary input are required to make the function reversible.

The value of the ancillary inputs can be chosen by the designer. Garbage outputs are by definition don't cares leading to an incompletely specified function. However, many synthesis approaches require a completely specified function so that all don't cares are assigned a concrete value. The adder is embedded in a reversible function including four variables,
one ancillary input, and two garbage outputs. 

\begin{example}
One possible assignment to the ancillary input and the don't care values of the garbage output are shown in Table~\ref{wille_table:emb_adder}.
Here, the target full adder is obtained if the ancillary input~c$_{in}$ is set to 0
and outputs c$_{out}$ and sum are observed.  $g_1$ and $g_2$ are the two garbage outputs.
\end{example}

State-of-the-art synthesis approaches are of two kinds: In \emph{functional synthesis},
the target function is embedded into a fully reversible function as sketched above (using methods such as in~\cite{DBLP:journals/corr/SoekenWKMD14,ZulehnerW17Emb}). The resulting function is synthesized using methods directly relying on reversible function descriptions~\cite{GWDD:2009, SWH+:2012, DBLP:journals/jsc/SoekenTDD16, DBLP:conf/rc/SoekenDM16}.
In \emph{structural synthesis}, the target function is first represented by dedicated function descriptions such as \emph{Exclusive Sum of Products}~(ESoPs)
or \emph{Binary Decision Diagrams}~(BDDs). They are then mapped to reversible circuits. Here, embedding is implicitly conducted during synthesis. ~\cite{FTR:2007,WDb:2009} are well-known structural synthesis approaches.
In both schemes, the use of ancillary inputs and garbage outputs
is essential to realize a non-reversible target
function in reversible logic.

\section{Reverse Engineering of Reversible Circuits}

Reversible circuits obtained using the state-of-the-art synthesis methods are challenging to an adversary\footnote{either in the foundry or an end user.} who has access to the gate level implementation aiming to reverse engineer
a circuit. We illustrate these challenges first. Then, we discuss several reverse engineering threat models.

\subsection{Challenges for Reverse Engineering}
{\bf An adversary who has access to the gate level implementation of a circuit realizing a reversible function can trivially reverse engineer it}. However, when a non-reversible function is embedded into a reversible circuit, a major challenge for the reverse engineer, who is unaware of the location of the inputs and outputs of the target function, is to identify the value and the location of the ancillary input bits and the location of the garbage output bits -- both are vital to reverse engineer the target function. 

\begin{example} Consider the full adder embedding shown in Fig.~\ref{fig:ex_adder}.
If the attacker is unaware of the location and values of the ancillary inputs, he/she cannot determine the target function. If $c_{in}$ is the ancillary input bit, setting $c_{in}$ to $0$, results in \mbox{$sum = x \oplus y$}, while setting $c_{in}$ to $1$, results in $sum = \overline{ x \oplus y}$.  
Another challenge for the attacker is to identify the primary and garbage output bits. 
In Fig.~\ref{fig:ex_adder}, any of the four output bits can be a garbage output. From the designer's perspective, the reversible circuit in Fig.~\ref{fig:ex_adder} realizes a full adder. Hence, he/she knows the \textit {primary outputs} and the \textit{garbage outputs} of the target function. If this information is unavailable, it is difficult to reverse engineer the functionality.
\end{example}

Overall, the ancillary inputs and garbage outputs can naturally hide the embedded, non-reversible target function. The key question is how can an attacker identify the ancillary inputs and determine their values and identify the garbage outputs from a gate-level netlist? 

In the remainder of this paper, we will discuss these reverse engineering challenges. Different threat models will be considered according to the level of knowledge the attacker has about the synthesis approach used to derive the circuit. More precisely, what if the attacker
\begin{itemize}
\item is ignorant of the synthesis method used to generate the circuit,
\item knows the synthesis method used to generate the circuit.
\end{itemize}

\subsection{Reverse Engineering without Knowing the Synthesis Approach}
\label{Attack Without Knowing the Synthesis Approach}
If the attacker does not know how the circuit has been obtained, he/she cannot distinguish the garbage outputs from the primary outputs and the ancillary inputs from the primary inputs. 
Hence, his/her primary goal is to determine the location and value of the ancillary inputs as well as the location of the garbage outputs. We first derive the number of possible target functions embedded into a reversible circuit. This presents the difficulty of reverse engineering for an attacker. This analysis is independent of the synthesis approach used to generate the reversible circuit.

Assume that a reversible circuit has $n$ input/output bits. There are $2^n$ possible input/output combinations. The circuit can be represented as a function $f(x_1,x_2,\cdots,x_n) = (y_1, y_2,\cdots, y_n)$. Each $x_i$ can be either a primary input of the target function
or an ancillary input, while each $y_i$, can be either a primary output of the target function or a garbage output, where $i$ varies from $1$ to $n$. 

Each output $y_i$, can be computed as $y_i= f_i(x_{i_1}, x_{i_2},\cdots, x_{i_{m_{i}}})$, where $m_i$ ($1\leq m_{i} \leq n$) is the number of inputs that drive $y_i$. Let $k_i$ ($0\leq k_{i} \leq m_{i}$) be the number of inputs that drive $y_i$ but not $y_p$ where $p$ anywhere in the interval from 1 to $i-1$ ($1\leq p < i$). The number of possible embedded functions in $y_i$ can be obtained by considering any subset of the $k_{i}$ inputs as ancillary inputs. Thus, the number of possible embedded functions for $y_i$ is
\begin{equation}
\begin{aligned}
\small e(k_i) =\sum_{j=0}^{k_i} C(k_i,j)\times2^j
\end{aligned}
\end{equation}
 where the binomial coefficient $C(k_i,j)$ refers to the number of ways (combinations) of selecting $j$ unordered ancillary inputs from $k_i$ inputs. We denote the number of embedded functions with $n$ primary output bits of a reversible circuit as
\begin{equation}
\begin{aligned}
{\small E(n,K)=\prod_{i=1}^n e(k_i)}
\end{aligned}
\end{equation}
where $k_{i}\in K$.

However, each output bit of a reversible circuit can be either a primary or a garbage output. Thus, the number of all possible target functions embedded into a reversible circuit with $n$ inputs/outputs is
\begin{equation}
\begin{aligned}
{\small EMB(n,K)= (2^{n}-1)\times E(n,K)}
\end{aligned}
\end{equation}
where $2^{n}-1$ indicates the number of all possible combinations of the output bits of a reversible circuit. We subtract one to exclude the case where all the output bits of a reversible circuit are garbage.

\begin{example} In Fig.~\ref{fig:rev}, $k_{1}=3, k_{2}=0,$ and $k_{3}=0$. Thus, $e_{1}=27, e_{2}=1,$ and $e_{3}=1$. The number of possible embedded functions is $27 \times (2^{3}-1) = 189$. \end{example}

From the attacker's perspective, it is hard to differentiate the target function from the other possible function embeddings. {\bf The number of possible embeddings can thus be used as the metric to assess the strength of defenses for reversible circuits against reverse engineering.}

\subsection{Reverse Engineering Knowing the Synthesis Approach}
\label{Attack with the Knowledge of the Synthesis Approach}

Next, we analyze how the attack changes when the adversary is aware of the synthesis approach used to generate the reversible circuit. More precisely, does the knowledge of the synthesis approach help the attacker to identify the positions and values of constant inputs and garbage outputs?

The attacker may utilize  visualization tools such as~\cite{DBLP:conf/rc/WilleSSDD14}  to highlight the structure and properties of a reversible circuit. Based on this information, the structure of the reversible circuit can be analyzed, the applied synthesis approach can be identified, and, possibly the target function can be reverse engineered. 

In the following, we discuss this case 
considering circuits obtained by both, structural synthesis and functional synthesis reviewed in Section~\ref{sec:syn_rev}. BDD-based synthesis~\cite{WDb:2009} represents a structural synthesis approach and QMDD-based synthesis~\cite{6165069} represents a functional synthesis approach.
\begin{figure}[htb] 
	\centering
	\includegraphics[width=3.2in]{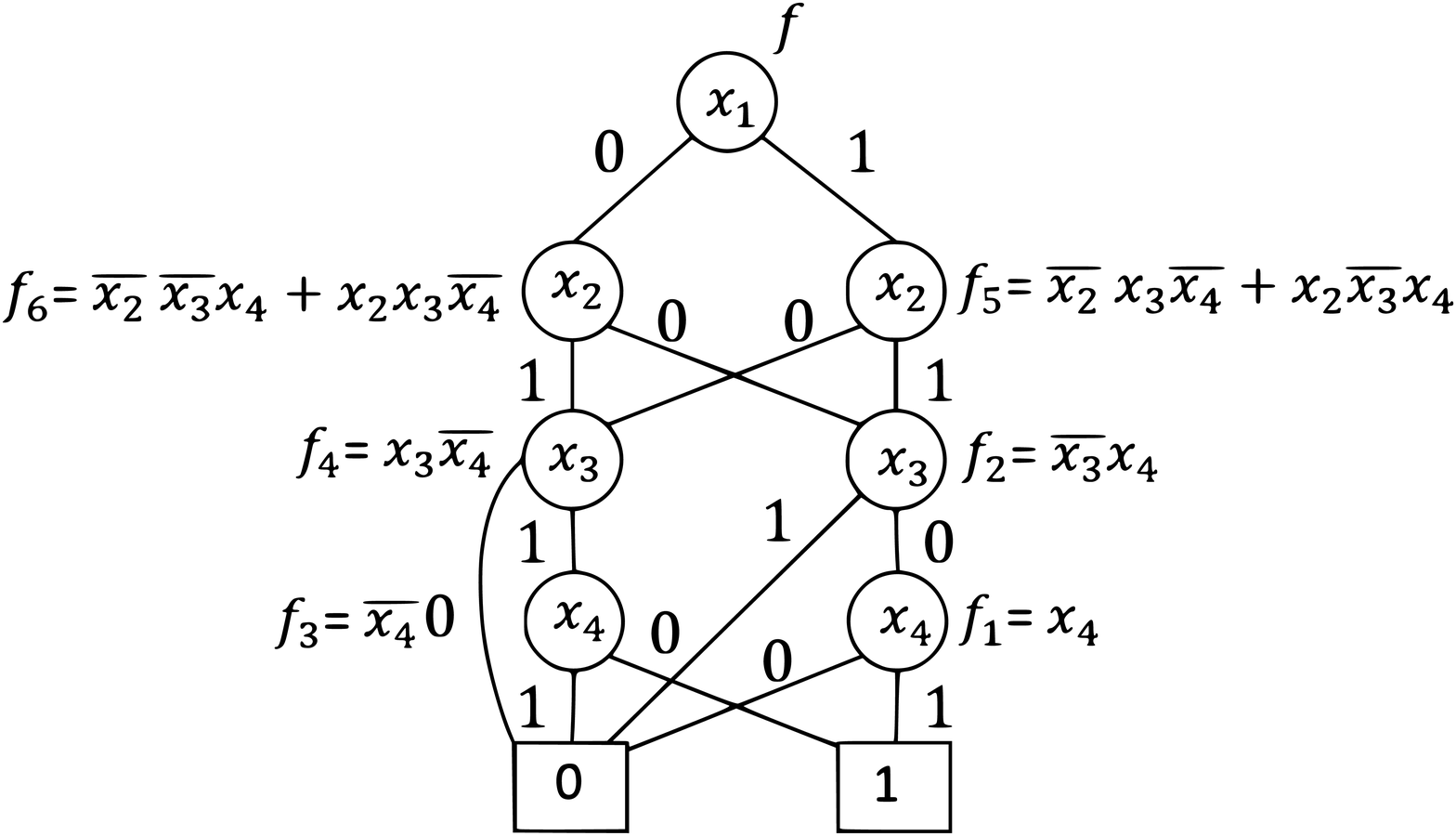}
		\caption{BDD of~$f=\overline{x}_1\overline{x}_2\overline{x}_3x_4+\overline{x}_1x_2x_3\overline{x}_4+x_1\overline{x}_2x_3\overline{x}_4+x_1x_2\overline{x}_3x_4$
	}\label{fig:dd}
\end{figure}
\subsubsection{BDD-based Synthesis of Reversible Logic} 
\label{BDD synthesis}
In BDD-based synthesis as proposed in~\cite{WDb:2009}, the target function is provided in terms of  a \emph{Binary Decision Diagram (BDD)}~\cite{Bry:86}, i.e.~a directed acyclic
graph $G = (V, E)$ where e.g.~a Shannon decomposition
$$f=\overline x_i\cdot f_{x_i=0} + x_i\cdot f_{x_i=1}$$
is carried out in each node $v\in V$ . The function~$f_{x_i=0}$ ($f_{x_i=1}$) is the negative (positive) 
co-factor of~$f$ obtained by assigning $x_i$ to~$0$ (1).
In the following, the node
representing~$f_{x_i=0}$ ($f_{x_i=1}$) is denoted by~${\rm low}(v)$ (${\rm high}(v)$),
while~$x_i$ is called the select variable. 

\begin{example} Fig.~\ref{fig:dd} shows a BDD representing the function~$f=\overline{x}_1\overline{x}_2\overline{x}_3x_4+\overline{x}_1x_2x_3\overline{x}_4+x_1\overline{x}_2x_3\overline{x}_4+x_1x_2\overline{x}_3x_4$ as well as the respective co-factors resulting from the application of the Shannon decomposition. \end{example}

Given a BDD~$G = (V, E)$ of a function, a corresponding reversible circuit can easily 
be derived. To this end, all nodes~$v\in V$ of~$G$ are traversed in a 
depth-first fashion and substituted with
a cascade of reversible gates. The respective cascade of gates
depends on the successors of the node~$v$. Fig.~\ref{fig:look_up} shows a look-up table that maps different nodes of the BDD to their corresponding reversible sub-circuits. 
Note that this often requires an additional
(ancillary) circuit line in order to realize the non-reversible decomposition employed in this node.
To obtain a reversible circuit realizing $f$ the entire BDD should be traversed.

\begin{example} Consider the BDD from Fig.~\ref{fig:dd}. 
The co-factor~$f_1$ can easily be represented by the primary input~$x_4$. Having the value of~$f_1$ available, the co-factor~$f_2$ can be realized by the first two gates depicted in Fig.~\ref{fig:dd_circ}\footnote{Note that an additional circuit line is added to preserve the values of~$x_4$ and $x_3$ which are still needed by the co-factors~$f_3$ and $f_4$, respectively.}. 
In this manner, respective sub-circuits can be added for all remaining co-factors until a circuit representing the overall function~$f$ results. The remaining steps are shown in Fig.~\ref{fig:dd_circ}.\end{example}

Thus, to realize the target function, decomposition is applied leading to smaller sub-functions for which existing building blocks 
can be applied. Then, the resulting sub-circuits can be composed to realize the overall function.
Overall, this realizes arbitrary (i.e.~not necessarily reversible) functions -- at the expense of a huge number of additional lines leading to many ancillary inputs and garbage outputs. 
\subsubsection{Reverse Engineering BDD-based Reversible Circuits}
\label{Attack Procedure on BDD-based Reversible Circuits}
BDD-based synthesis yields structured reversible circuits from which several properties can easily be obtained. 
 
\begin{figure}[htbp] 
	\centering
	\includegraphics[width=2.7in]{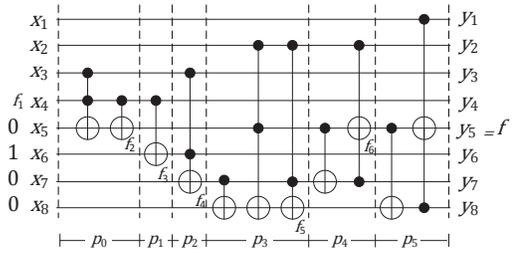}
	\caption{Reversible Circuit derived from the BDD of $f$, where $p_i$ is the $i_{th}$ partition of the reversible circuit.}
	\label{fig:dd_circ}
\end{figure}

\begin{figure}[htbp]
\centering
\subfigure []{\label{fig:obf1a}\includegraphics[width=1.8in]{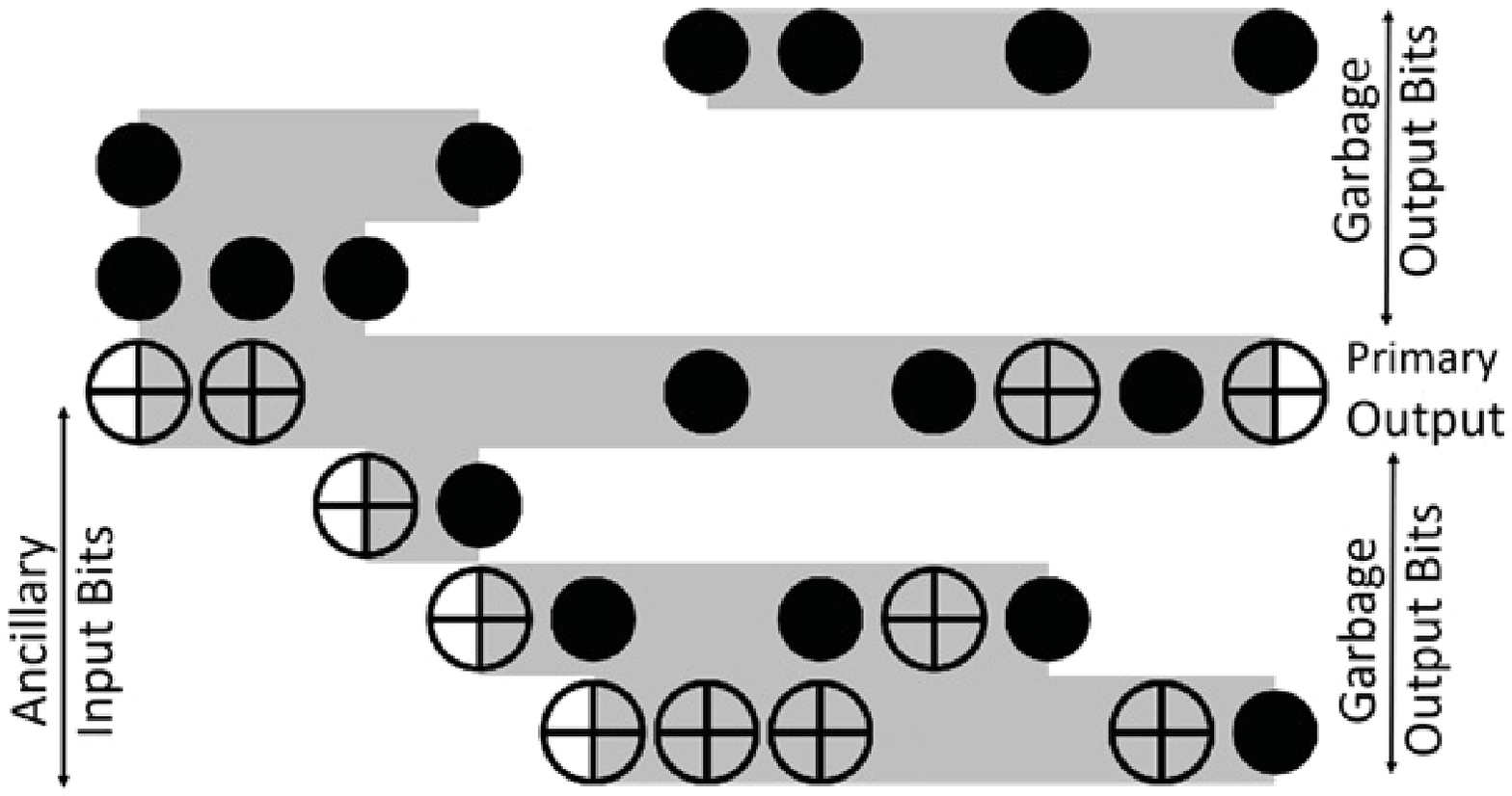}}\hspace{0.5em}
\subfigure []{\label{fig:obf1b}\includegraphics[width=1.6in]{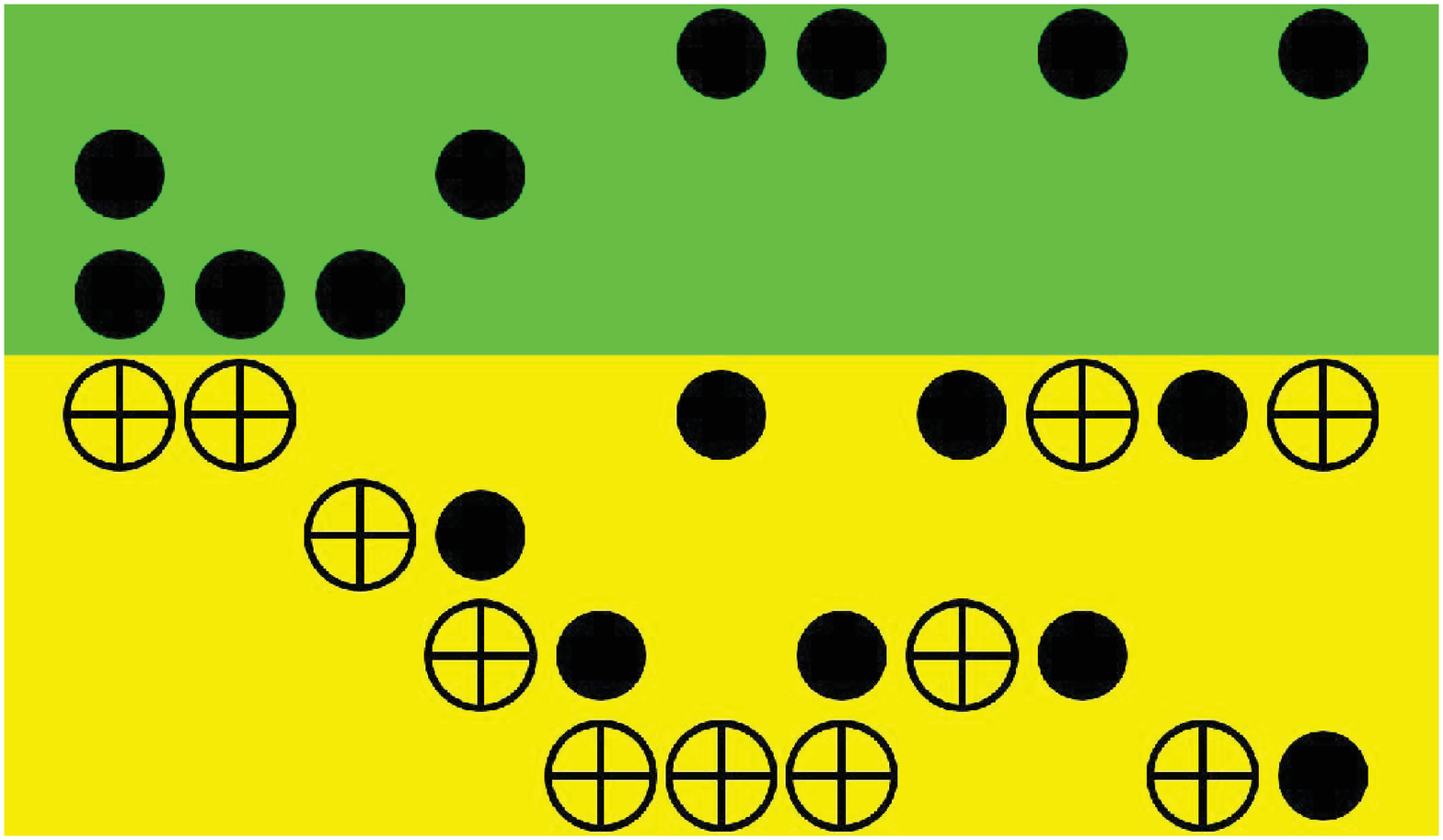}}               
\caption{Visualization of Reversible Circuits: (a) Ancillary Inputs and Garbage Outputs, (b) Target and Control Lines}
\label{fig:obf1}
\end{figure}

\begin{example} Consider the circuit obtained by BDD-based synthesis as shown in Fig.~\ref{fig:dd_circ}. Fig.~\ref{fig:obf1a} and Fig.~\ref{fig:obf1b} show a visualization (obtained by RevVis~\cite{DBLP:conf/rc/WilleSSDD14}) highlighting the positions of the ancillary inputs and garbage outputs
as well as the positions of control (green color) and target (yellow color) lines, respectively. The green region indicates the location of the primary inputs, which directly drive the garbage output bits of the BDD-based reversible circuit. The remaining inputs are ancillary inputs as illustrated in Fig.~\ref{fig:obf1a}. The primary output of the reversible circuit is connected to the target line of right most gate as shown in Fig.~\ref{fig:obf1a}. 
\end{example}
\begin{figure}[t]
\centering

\includegraphics[width=3in]{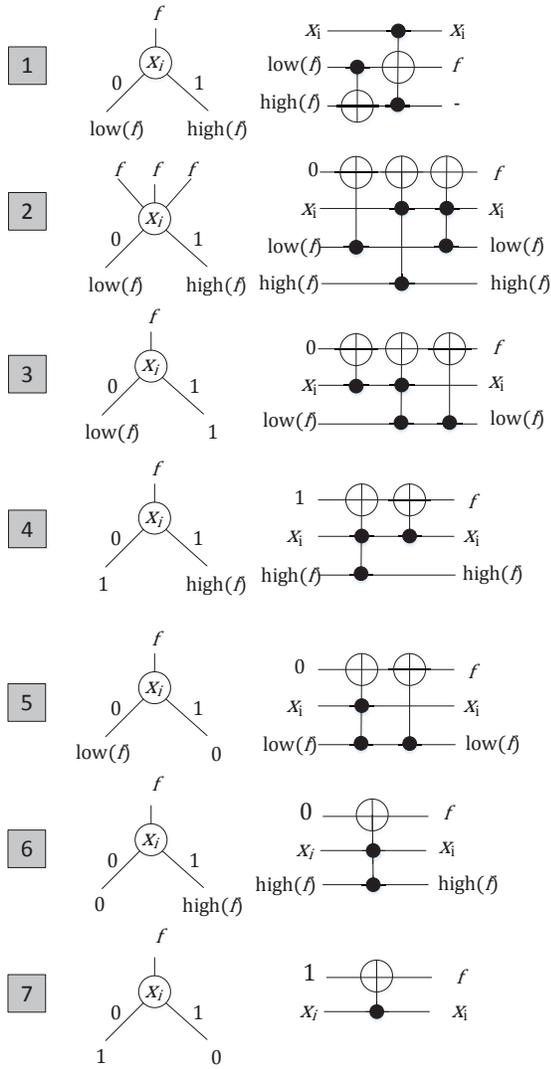}
\caption{Look-up table for reversible cascades representing Shannon decomposition of BDD. High($f$) (low($f$)) indicates the value of function $f$ when the input $x_i$ is set to 1 (0).}
\label{fig:look_up}
\end{figure} 
In circuits obtained using BDD-based synthesis,
\begin{enumerate}
\item the primary inputs are directly connected to the garbage outputs and
\item an intermediate output of a sub-circuit that has no control over other reversible gates is a primary output. 
\end{enumerate}

The attacker can exploit these properties to reveal the function embedded in a reversible circuit by:
\begin{enumerate}
\item Distinguishing between primary and garbage outputs.
\item Identifying the location of the ancillary inputs.
\item Identifying the value of the ancillary inputs. 
\end{enumerate}

First, the attacker distinguishes the primary and garbage outputs of the reversible circuit. As shown in Fig.~\ref{fig:obf1b}, the lines of the reversible circuit in the green region do not include any target line, and thus, they are connected to the garbage outputs, while the gate at the rightmost part of the reversible circuit is connected to the primary output.  
To this end, the second property discussed above can be exploited. 
Then, the attacker reveals the location of the ancillary inputs based on the first property. 
Finally, the attacker discovers the value of the ancillary inputs as follows:
\begin{enumerate}
\item Partitioning the reversible circuit into sub-circuits.
\item Identifying the ancillary input values of the sub-circuits.
\end{enumerate}
Decomposition of a target function generates different co-factors, which can be represented as nodes of the BDD. Each node is substituted with the corresponding sub-circuit. Some sub-circuits have a unique structure, which enables identifying the associated ancillary input value, while others behave as \emph{universal gates}, which can be reconfigured to support different sub-functions depending on the associated ancillary input value. A reverse engineer can extract the value of the ancillary inputs associated with sub-circuits which have a unique structure. Assuming that the attacker knows the decomposition used to generate the BDD-based reversible circuit, he/she can generate the look-up table shown in Fig.~\ref{fig:look_up}. The attacker partitions the reversible circuit into sub-circuits, where each sub-circuit consists of the maximum number of adjacent reversible gates that match at least one sub-circuit in the look-up table\footnote{This partitioning will maintain the structure of the BDD.}. Then the attacker maps each sub-circuit of the reversible circuit to the corresponding node in the look-up table to identify the associated ancillary input value, and thus, the target function.

\begin{example} In Fig.~\ref{fig:look_up}, case 4 and 5 show an example of a sub-circuit that can represent any of two co-factors (nodes) depending on the associated ancillary input value. On the other hand, the sub-circuits in case 1, 2, 3, 6, and 7 represent unique co-factors (nodes). \end{example}

\begin{example} To illustrate the attack let us consider the reversible circuit in Fig.~\ref{fig:dd_circ}. BDD-based synthesis using Shannon decomposition is utilized to create the reversible circuit. The attacker first identifies the primary output, which is $y_5$ using the second property of the BDD-based reversible circuit. Also, he/she identifies the location of garbage outputs as they are directly connected to the inputs; Here they are $y_1$, $y_2$, $y_3$, and $y_4$ derived using the first and second properties of the BDD-based reversible circuit. From the attacker perspective $y_6$, $y_7$, and $y_8$ are potential primary outputs. Then, the attacker determines the location of the ancillary inputs, which are $x_5$, $x_6$, $x_7$, and $x_8$ using the first property of the BDD-based reversible circuit. Next, the attacker partitions the reversible gates into sub-circuits as explained above. As shown in Fig.~\ref{fig:dd_circ}, the number of partitions is six. Lets consider $p_i$ as the $i^{th}$ partition of the reversible circuit. $p_1$, $p_2$, and $p_3$ are mapped to case 7, 6, and 2 in Fig.~\ref{fig:look_up}, respectively. Thus, $p_1$, $p_2$, and $p_3$ can be uniquely identified. As a result, the value of the ancillary inputs $x_6$, $x_7$, and $x_8$ are 1, 0, and 0, respectively. However, $p_0$ can be mapped to case 4 or 5. Each of these two cases results in a different co-factor, which requires a different ancillary input bit value. As a result, the number of possible functions at $y_5$ is $2$. In general, for $m$ unknown ancillary input bits that drive a primary output $y_i$, the number of possible realized functions at $y_i$ is $2^m$. For each of the remaining output bits $y_6$, $y_7$, and $y_8$, the number of possible embeddings is 1. Thus, the number of possible embeddings for the reversible circuit in Fig.~\ref{fig:dd_circ} is $2^3 \times 2$, in which $2^3$ indicates the number of all possible combinations of selecting potential primary output bits as primary output bits of the reversible circuit.
  \end{example}

{\bf As we show in the experimental section, one can successfully identify all the primary outputs of reversible circuits synthesized using BDD-based tools}. 

\subsubsection{Functional Synthesis of Reversible Logic} 
The embedded function in a reversible circuit may be hidden under different configurations of ancillary inputs and garbage outputs bits. Embedding a function results in a different number and value of ancillary inputs, while maintaining reversibility. In addition, the value of the garbage outputs as well as the outputs of the non-functional input assignments can be chosen arbitrary as long as they satisfy reversibility. Selecting the number and the value of the ancillary inputs combined with the possible locations where the garbage and the primary outputs can be placed can make reverse engineering of the target function more difficult. 

\begin{example} Table~\ref{wille_table:emb_adder} illustrates one possible embedding of the 1-bit adder into a reversible circuit. The value of the ancillary input that activates the target function, 0 in this example, can be replaced by a 1. Furthermore, one can assign different values to garbage output bits $g_1$ and $g_2$, while maintaining reversibility. For example, for input vector $c_{in}xy$ = 000,  $g_1$ $g_2$ can be assigned 00, 01, 10, or 11. The output of the non-functional input assignments, when the ancillary input is 1 in this example, such as $c_{in}xy$ = 000, can be assigned an arbitrary value as well, such as $c_{out}sum_{}g_1g_2$ = 1100, as long as the given output is not used as an output for a functional input assignment that precedes this output in Table~\ref{wille_table:emb_adder}.\end{example}

\mbox{Quantum Multiple-valued Decision Diagrams (QMDD)}-based synthesis~\cite{6165069} takes the reversible function generated by the embedding step and creates a reversible circuit. We omit the details of QMDD-based synthesis since the ancillary inputs and garbage outputs are placed during the embedding phase which is a preprocessing step prior to the functional synthesis approach. As discussed above, due to the embedding step, the structure of the resulting reversible circuit varies in the value, the number, and the location of the ancillary inputs as well as in the value and position of the garbage outputs.

\begin{figure}[htbp]
\centering
\subfigure []{\label{fig:Rev_QMDD1}\includegraphics[width=3.5in]{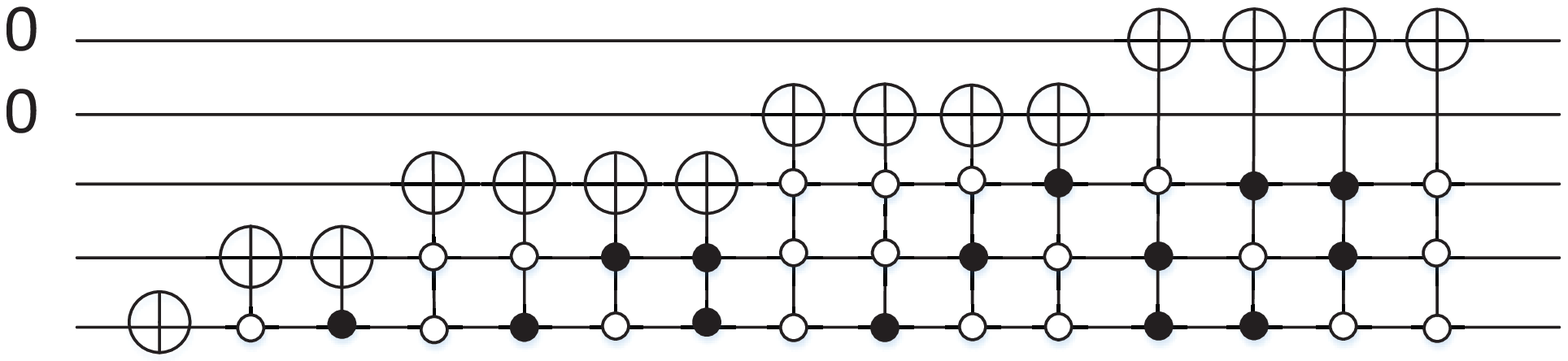}}\\ 
\subfigure []{\label{fig:Rev_QMDD2}\includegraphics[width=1.4in]{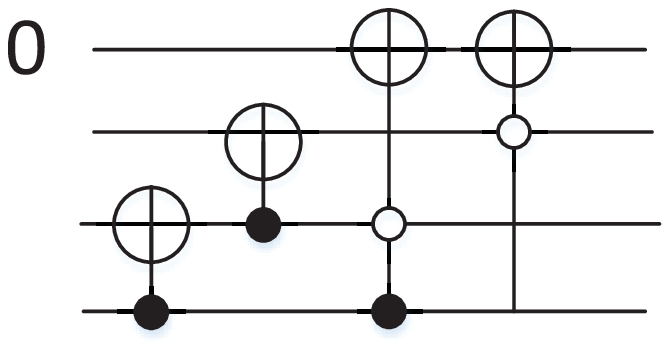}}               
\subfigure []{\label{fig:Rev_QMDD3}\includegraphics[width=1.4in]{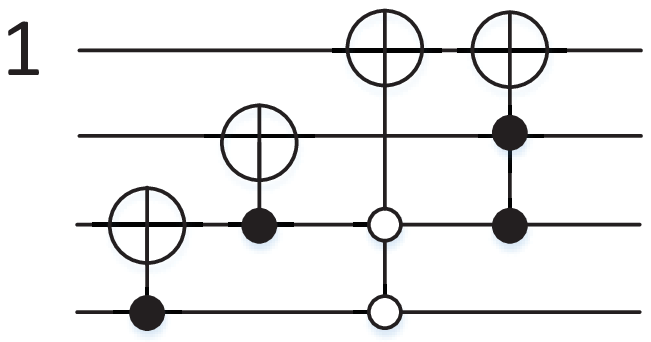}}               
\caption{QMDD-based reversible circuit of a 1-bit adder with: (a) two ancillary input bits of value 00, (b) a single ancillary input bit of value 0, (c) a single ancillary input bit of value 1.}
\label{fig:Rev_QMDD}
\end{figure}

\begin{example} Fig.~\ref{fig:Rev_QMDD} shows the three possible reversible circuits for a 1-bit adder shown in Table~\ref{wille_table:adder} which have been obtained by QMDD-based synthesis with different values and numbers of ancillary inputs, garbage outputs, and outputs of non-functional input assignments. \end{example}

Assume that the attacker is unaware of the number, the value, and the location of ancillary input bits as well as the number and the value of the garbage outputs. That means, even if the attacker knows the used synthesis scheme, the number of possible target functions remains the same. The complexity of reverse engineering is introduced by the embedding phase and not the synthesis step, which makes it difficult to reverse engineer a reversible circuit by a rogue in the foundry even if he/she has access to the gate level netlist of the synthesized reversible circuit with the knowledge of the applied functional synthesis approach.

\section{Input and Output Scrambling}
We have shown that the functional synthesis approaches (such as QMDD) are more resilient than structural synthesis approaches (such as BDD) to reverse engineering. To hamper reverse engineering of BDD-based reversible circuits, we propose scrambling the inputs or outputs of a synthesized reversible circuit by adding extra ancillary inputs or garbage outputs. Scrambling of inputs entails, not identifying which inputs are primary inputs and which inputs are ancillary inputs and what constant values are assigned to these ancillary inputs, while scrambling of the output entails, not identifying which outputs are primary outputs and which are garbage outputs. 

 Extra ancillary inputs or garbage outputs are added to the target function before the synthesis step. While the knowledge of the structural synthesis approach can assist in identifying the location and the value of the ancillary inputs and the location of the garbage output, the extra ancillary inputs and garbage outputs are independent of the structural synthesis approach. Scrambling the ancillary inputs and the garbage outputs will hinder an attacker's ability to discover the target function.
 
Structural synthesis approaches, which realize arbitrary functions add additional ancillary inputs and garbage outputs regardless of the target functions, and thus, resulting in large number of 
circuit lines but relatively small quantum cost. While the attacker is able to identify the majority of the ancillary input bits of a reversible circuit generated using BDD-synthesis approach, in the presence of the extra ancillary inputs added prior to the realization of the embedded function, every input is a potential ancillary input. As a result, the number of possible embeddings increases significantly. 

\begin{figure}[htbp]
\centering
\subfigure []{\label{fig:bdd_c}\includegraphics[width=3.5in]{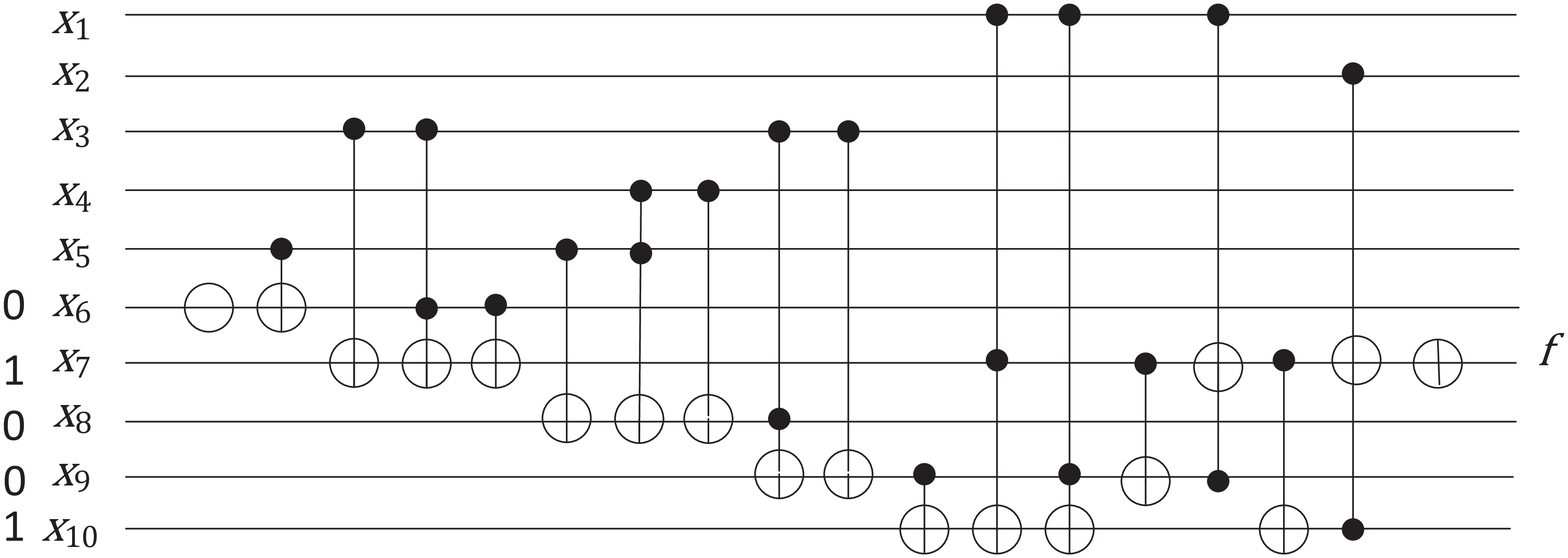}}\\ 
\subfigure []{\label{fig:bdd_g}\includegraphics[width=2.9in]{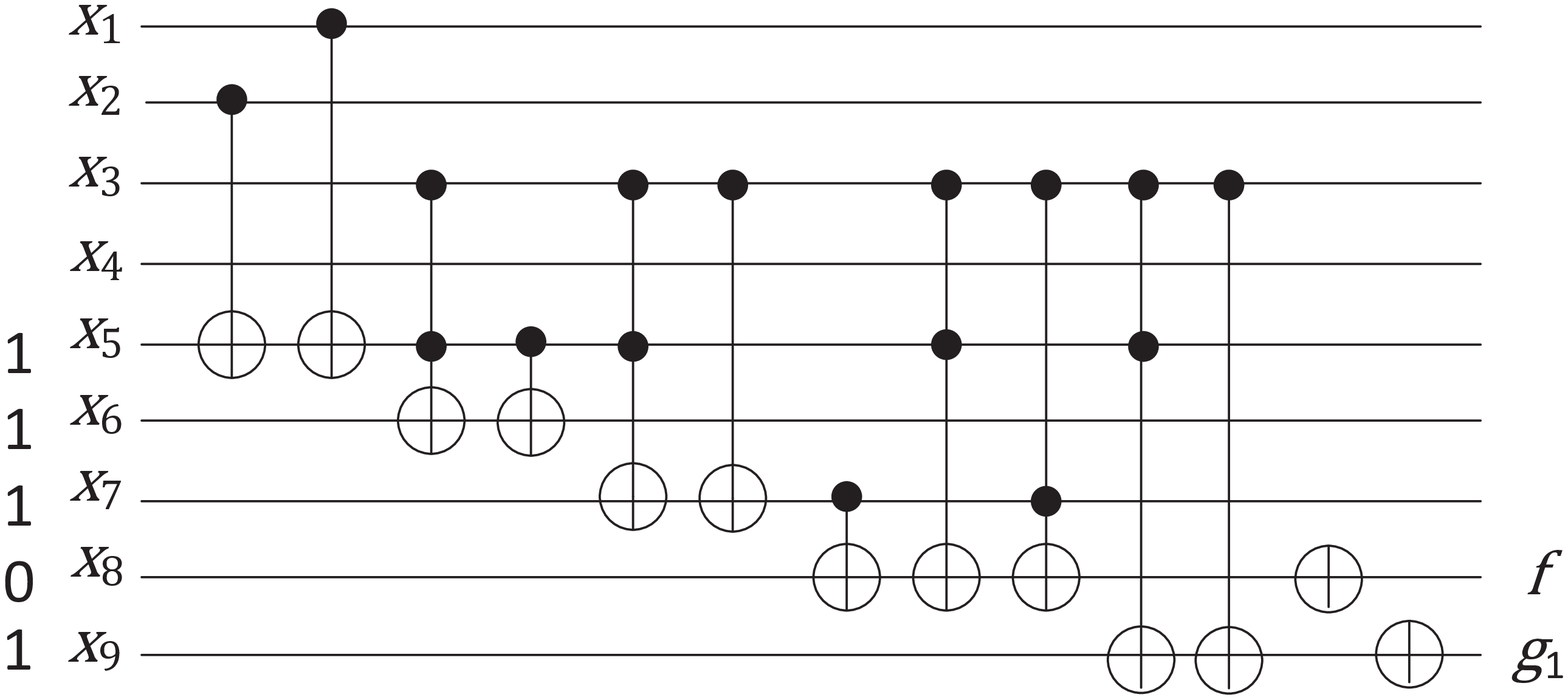}}               
\caption{The reversible circuit of the target function in Fig.~\ref{fig:dd} in the presence of (a) single extra ancillary input and (b) single extra garbage output. }
\label{fig: bdd_c_g}
\end{figure}

\begin{example}
Fig.~\ref{fig:bdd_c} illustrates the BDD-based reversible circuit of function $f$ in Fig.~\ref{fig:dd} in the presence of an extra ancillary input $x_5$ of value $0$. Our proposed attack with the information of the synthesis approach in Section~\ref{Attack Procedure on BDD-based Reversible Circuits} can identify three ancillary input bits out of five introduced by the synthesis approach. However, the extra ancillary input can be any input classified as primary input by a reverse engineer. $x_{1}, x_{2}, x_{3}, x_{4},$ or $x_{5}$ are possible locations for the ancillary input.
\end{example}

Moreover, in the presence of extra garbage bits, every output of a reversible circuit generated using BDD-based (and thus structural) synthesis can be a potential output bit, which violates the second property of BDD-based reversible circuits as indicated in Section~\ref{Attack Procedure on BDD-based Reversible Circuits}. Thus, although the attacker can identify most of the ancillary input bits, he/she can no longer differentiate the primary outputs from the garbage outputs, resulting in a significant number of possible embeddings.

\begin{example}
Fig.~\ref{fig:bdd_g} illustrates the BDD-based reversible circuit of function $f$ in Fig.~\ref{fig:dd} in the presence of an extra garbage output $g_1$. Both $g_1$ and $f$ are connected to intermediate outputs of sub-circuits that do not control other reversible gates. In other words, the attacker can no longer distinguish $g_1$ from $f$.
\end{example}

The size of the additional garbage outputs and ancillary inputs is driven by two optimization criteria, (1) maximizing the number of embeddings ,and thus, the security level, and (2) minimizing the quantum cost. Increasing the number of garbage outputs and ancillary inputs results in increasing not only the number circuit lines and the quantum cost, but also the number of embeddings. In the experimental section, we analyze the trade-off between the security level in the presence of extra ancillary inputs/garbage outputs and their associated quantum cost.

Functional synthesis approaches can exploit the extra ancillary inputs and garbage outputs effectively to support reversibility at the expensive of large quantum cost.
\begin{example} An ancillary input $k$ can be added to the target function in Table~\ref{wille_table:adder} with a value 0. The embedding phase takes updated target function as an input and generates the corresponding reversible function. One potential reversible function is given in Table~\ref{wille_table:emb_adder}. 
 Similarly, one can insert a garbage output bit such as $g_1$ to the target function in Table~\ref{wille_table:adder}, which can also lead to the reversible function in Table~\ref{wille_table:emb_adder}.\end{example}

\begin{tiny}
\begin{table*}[!htbp]
\begin{center}
\caption{Number of embedded functions in reversible circuits created using BDD-based synthesis.}
\label{tab:R1}
\begin{tabular}{|l|c|c|c|c|c|c|c|c|}
\hline
{\multirow{2}{*}{Benchmark}} &\multicolumn{3}{c|}{\#} & Quant. & \multicolumn{2}{l|}{Without synthesis Info.} & \multicolumn{2}{l|}{With synthesis info.} \\\cline{2-4} \cline{6-9}
 &I/O&Garbage &Ancilla & Cost & \% & \#Embed & \% & \#Embed \\
  & & & & & Garbage & func. & D\_Ancill & func. \\\hline
4mod5\_8 & 7 & 6 & 3 & 24 & 66.7 & 15309 & 66.7 & 8 \\
4mod7\_26 & 12 & 10 & 8 & 86 & 40 & 1.36E+08 & 100 & 32 \\
4\_49\_7 & 15 & 11 & 11 & 126 & 36.4 & 2.94E+10 & 81.8 & 512 \\
5xp1\_90 & 30 & 20 & 23 & 254 & 30 & 3.45E+21 & 91.3 & 65536 \\
9symml\_91 & 27 & 26 & 18 & 206 & 34.6 & 2.00E+18 & 100 & 131072 \\
add6\_92 & 54 & 47 & 42 & 499 & 25.5 & 2.56E+38 & 100 & 3.44E+10 \\
adr4\_93 & 16 & 11 & 8 & 74 & 72.7 & 1.10E+10 & 100 & 8 \\
aj-e11\_81 & 16 & 12 & 12 & 130 & 33.3 & 1.76E+11 & 83.3 & 1024 \\
alu1\_94 & 28 & 20 & 16 & 139 & 60 & 1.50E+18 & 93.8 & 512 \\
alu2\_96 & 105 & 99 & 95 & 1436 & 10.1 & 2.69E+76 & 89.5 & 6.34E+29 \\
alu3\_97 & 66 & 58 & 56 & 644 & 17.2 & 2.23E+48 & 76.8 & 2.31E+18 \\
alu4\_98 & 541 & 533 & 527 & 7222 & 2.6 & 9.27169e+376 & 93.4 & 5.90E+166 \\
apex2\_101 & 498 & 495 & 459 & 5922 & 7.9 & 1.1339e+317 & 96.3 & 2.44E+142 \\
apex4\_103 & 547 & 528 & 538 & 8343 & 1.7 & 1.24077e+382 & 91.4 & 1.21E+170 \\
apla\_107 & 103 & 91 & 93 & 1002 & 11 & 1.38E+77 & 82.8 & 1.58E+29 \\
sao2\_199 & 74 & 70 & 64 & 667 & 14.3 & 3.74E+54 & 73.4 & 1.51E+23 \\
C7552\_119 & 35 & 19 & 30 & 202 & 26.3 & 5.37E+25 & 50 & 5.37E+08 \\
clip\_124 & 66 & 61 & 57 & 704 & 14.8 & 4.45E+48 & 78.9 & 1.84E+19 \\
cm150a\_128 & 37 & 36 & 16 & 186 & 58.3 & 2.95E+22 & 0 & 2.15E+09 \\
cm151a\_129 & 49 & 40 & 30 & 298 & 47.5 & 2.57E+32 & 70 & 1.07E+09 \\
cm42a\_125 & 22 & 12 & 18 & 117 & 33.3 & 8.23E+15 & 50 & 131072 \\
cm82a\_126 & 13 & 10 & 8 & 82 & 50 & 4.07E+08 & 100 & 32 \\
cm85a\_127 & 36 & 33 & 25 & 275 & 33.3 & 5.04E+24 & 56 & 8.59E+09 \\
cmb\_134 & 43 & 40 & 27 & 158 & 40 & 4.41E+28 & 100 & 1.68E+07 \\
co14\_135 & 27 & 26 & 13 & 159 & 53.8 & 6.25E+16 & 100 & 4096 \\
con1\_136 & 16 & 14 & 9 & 96 & 50 & 2.20E+10 & 77.8 & 512 \\
cordic\_138 & 52 & 50 & 29 & 325 & 46 & 3.47E+33 & 79.3 & 8.59E+09 \\
cu\_141 & 38 & 28 & 24 & 220 & 50 & 2.27E+25 & 66.7 & 4.19E+06 \\
dc1\_142 & 20 & 13 & 16 & 160 & 30.8 & 2.29E+14 & 87.5 & 2048 \\
decod\_137 & 35 & 19 & 30 & 202 & 26.3 & 5.37E+25 & 50 & 5.37E+08 \\
dist\_144 & 79 & 74 & 71 & 975 & 10.8 & 1.16E+59 & 100 & 7.38E+19 \\
dk17\_145 & 58 & 47 & 48 & 426 & 21.3 & 1.33E+42 & 77.1 & 2.81E+14 \\
ex5p\_154 & 206 & 143 & 198 & 1843 & 5.6 & 7.78E+157 & 65.7 & 1.29E+61 \\
example2\_156 & 105 & 99 & 95 & 1436 & 10.1 & 2.69E+76 & 89.5 & 6.34E+29 \\
f2\_158 & 16 & 12 & 12 & 113 & 33.3 & 1.76E+11 & 58.3 & 8192 \\
f51m\_159 & 385 & 377 & 371 & 5392 & 3.7 & 1.43E+252 & 91.9 & 2.02E+118 \\
hwb6\_14 & 46 & 40 & 40 & 507 & 15 & 9.74E+33 & 92.5 & 1.37E+11 \\
hwb7\_15 & 73 & 67 & 66 & 909 & 10.4 & 4.99E+54 & 92.4 & 1.84E+19 \\
hwb8\_64 & 112 & 105 & 104 & 1461 & 7.6 & 1.29E+80 & 95.2 & 2.54E+30 \\
max46\_177 & 54 & 53 & 45 & 598 & 17 & 2.05E+39 & 77.8 & 1.80E+16 \\
misex3\_180 & 428 & 414 & 414 & 4661 & 3.4 & 8.80261e+312 & 68.6 & 3.51E+159 \\
mlp4\_184 & 103 & 95 & 95 & 1158 & 8.4 & 5.51E+77 & 94.7 & 4.95E+27 \\
sqn\_203 & 40 & 37 & 33 & 426 & 18.9 & 1.04E+29 & 69.7 & 1.10E+12 \\
sqrt8\_205 & 30 & 26 & 22 & 240 & 30.8 & 8.64E+20 & 72.7 & 1.68E+07 \\
xor5\_195 & 6 & 5 & 1 & 8 & 100 & 729 & 100 & 1 \\
z4ml\_225 & 14 & 10 & 7 & 66 & 70 & 6.07E+08 & 100 & 8 \\
tial\_214 & 578 & 570 & 564 & 7609 & 2.5 & 9.75474e+379 & 94.9 & 1.27E+176 \\
urf2\_73 & 209 & 201 & 201 & 3187 & 4 & 6.33E+145 & 89.1 & 5.27E+64
 \\\hline
\end{tabular}
\end{center}
\end{table*}
\end{tiny}

\section{Experimental Results}
To evaluate the difficulty of extracting the target (non-reversible) function from the reversible circuit, we have conducted various
experiments. In each experiment, we report the number of possible embeddings as a security metric and the quantum cost. We show the trade off between the security level and the hardware cost. First, we reverse engineer reversible circuits without the knowledge of the synthesis approach that generates the reversible circuit. Second, we reverse engineer reversible circuits with the knowledge of the applied synthesis approach. Finally, we launch our attacks on reversible circuits generated after adding extra ancillary inputs or garbage outputs to the target function. We consider reversible circuits generated using BDD-based synthesis as an example of the structural synthesis and QMDD-based synthesis as an example of the functional synthesis provided in RevLib~\cite{4539430}. 

\subsection{Reverse Engineering without Knowing the Synthesis Approach}
In the first set of the experiments we assume that the attacker does not know the synthesis approach used to generate the reversible circuit. In Table~\ref{tab:R1} and ~\ref{tab:R2}, we measure the number of possible embeddings of BDD- and QMDD-based reversible circuits, respectively. In Table~\ref{tab:R1}, column 1 through 5 indicate the target circuit name, the number of input/output bits, garbage output bits, and ancillary input bits, and the quantum cost of the reversible circuit, respectively. Column 6 and 7 are measured under the assumption that the attacker does not know the applied synthesis approach. \%Garbage in column 6 reports the percentage of leaked garbage output bits due to the direct connection between some of the inputs and outputs of the reversible circuit. These output bits can be easily classified as garbage outputs if the attacker has access to the gate level implementation of the circuit. 
\#Embed func. reports the number of possible embeddings in column 7.

Table~\ref{tab:R1} shows that on average the attacker can identify 42\% of the garbage output bits in BDD-based reversible circuits. However, the number of garbage output bits of BDD-based reversible circuits is significantly large. This is due to the fact that every node of the BDD can produce a garbage output. 
Moreover, the location and the value of the ancillary inputs are unknown to the attacker forcing him to explore a large number of possible embeddings using equation 3 in Section~\ref{Attack Without Knowing the Synthesis Approach}.

The QMDD-based synthesis approach generates the reversible circuits whose core data is summarized in Table~\ref{tab:R2}, using arbitrary values to ancillary inputs that satisfy the reversibility. 
Column 1 through 4 indicate the circuit name, the number of input/output bits, and garbage output bits, and the percentage of leaked garbage output bits by the attacker due to the direct connection between inputs and outputs of the reversible circuit, respectively. Columns 5 and 6 summarize the quantum cost and the number of possible embeddings, respectively. 

The results show that the percentage of identified garbage output bits of QMDD-based reversible circuits is 35\% on average. The number of embeddings of reversible circuits generated using BDD and QMDD synthesis approach is very large. However, the quantum cost of QMDD-based reversible circuits exceeds the corresponding one of BDD-based reversible circuits. We conclude that while reverse engineering reversible circuits with no information of the applied synthesis approach is always difficult, the quantum cost varies significantly based on the applied synthesis approach. 

\begin{tiny}
\begin{table}[htbp]
\centering
\caption{Number of embedded functions in reversible circuits created using QMDD-based synthesis.}
\label{tab:R2}
\begin{tabular}{|c|c|c|c|c|c|} \hline
\multirow{2}{*}{Benchmark}     & \#Inp & \multicolumn{2}{l|}{Garbage} &  Quant. & \#Embed  \\\cline{3-4}
        &      /Out     & \#  & \% &         Cost   &  Cir             \\\hline
4mod5-v0\_18 & 5 & 4 & 0.0 & 93 & 5.49E+06 \\
4mod7-v1\_96 & 5 & 3 & 33.3 & 187 & 2.31E+06 \\
5xp1\_194 & 17 & 7 & 14.3 & 2803 & 6.75E+35 \\
add6\_196 & 19 & 12 & 8.3 & 16626 & 5.91E+39 \\
adr4\_197 & 13 & 8 & 12.5 & 1625 & 4.98E+22 \\
aj-e11\_165 & 4 & 0 & - & 89 & 710775 \\
alu1\_198 & 20 & 12 & 25.0 & 215 & 5.65E+46 \\
alu2\_199 & 16 & 10 & 40.0 & 28010 & 2.14E+30 \\
alu3\_200 & 18 & 10 & 0.0 & 11156 & 2.54E+47 \\
alu4\_201 & 22 & 14 & 28.6 & 3987932 & 2.39E+58 \\
apex4\_202 & 28 & 9 & 44.4 & 282265 & 9.32E+95 \\
apla\_203 & 22 & 10 & 0.0 & 18935 & 3.77E+80 \\
C7552\_205 & 21 & 5 & 40.0 & 980 & 1.77E+74 \\
clip\_206 & 14 & 9 & 44.4 & 35353 & 1.45E+31 \\
cm150a\_210 & 22 & 21 & 0.0 & 2023 & 2.13E+75 \\
cm151a\_211 & 28 & 19 & 68.4 & 8415 & 2.32E+49 \\
cm152a\_212 & 12 & 11 & 72.7 & 236 & 5.19E+10 \\
cm42a\_207 & 14 & 4 & 0.0 & 307 & 1.51E+36 \\
cm82a\_208 & 8 & 5 & 40.0 & 155 & 7.32E+10 \\
cm85a\_209 & 14 & 11 & 45.5 & 10242 & 2.31E+32 \\
cmb\_214 & 20 & 17 & 23.5 & 32775 & 8.24E+49 \\
co14\_215 & 15 & 14 & 14.3 & 229375 & 1.07E+25 \\
con1\_216 & 9 & 7 & 42.9 & 254 & 1.20E+13 \\
cordic\_218 & 25 & 23 & 8.7 & 102742609 & 7.00E+39 \\
cu\_219 & 25 & 15 & 33.3 & 9674 & 1.47E+67 \\
dc1\_220 & 11 & 4 & 0.0 & 368 & 6.72E+23 \\
dk17\_224 & 21 & 10 & 0.0 & 8284 & 8.62E+76 \\
example2\_231 & 16 & 10 & 40.0 & 28010 & 3.04E+36 \\
f51m\_233 & 22 & 14 & 42.9 & 1028459 & 1.83E+66 \\
hwb6\_56 & 6 & 0 & - & 7358 & 6.01E+11 \\
hwb7\_59 & 7 & 1 & 0.0 & 31411 & 2.74E+15 \\
hwb8\_113 & 8 & 1 & 0.0 & 150218 & 3.68E+19 \\
misex3\_242 & 28 & 14 & 50.0 & 3321686 & 3.25E+122 \\
mlp4\_245 & 16 & 8 & 75.0 & 12345 & 5.72E+47 \\
sqn\_258 & 10 & 7 & 0.0 & 3225 & 4.61E+20 \\
sqrt8\_260 & 12 & 8 & 0.0 & 1529 & 8.82E+27 \\
xor5\_254 & 6 & 5 & 60.0 & 16 & 2.18E+07 \\
urf2\_277 & 8 & 0 & - & 120705 & 3.68E+19 \\
table3\_264 & 28 & 14 & 35.7 & 4310154 & 1.51E+149 \\
sao2\_257 & 14 & 10 & 0.0 & 47224 & 6.94E+32 \\
pm1\_249 & 14 & 4 & 0.0 & 307 & 2.99E+40 \\
radd\_250 & 13 & 8 & 0.0 & 1392 & 1.15E+32 \\
root\_255 & 13 & 8 & 25.0 & 11798 & 1.04E+33 \\
ryy6\_256 & 17 & 16 & 87.5 & 254477 & 2.90E+26 \\
sym9\_317 & 27 & 26 & 34.6 & 762136380 & 3.05E+155 \\
squar5\_261 & 13 & 5 & 20.0 & 507 & 1.26E+35 \\
x2\_267 & 17 & 10 & 30.0 & 1339 & 2.31E+32 \\
\hline
\end{tabular}
\end{table}
\end{tiny}

\subsection{Reverse Engineering Knowing the Synthesis Approach}
In the second set of the experiments, we assume that the attacker knows the synthesis approach used to generate the reversible circuit. We launch our attack on BDD-based reversible circuits, while we note that reverse engineering QMDD-based reversible circuits with/without knowing the synthesis approach yields the same results. Column 8 through 9 in Table~\ref{tab:R1} provide the percentage of recovered ancillary input bits and the number of possible embeddings after reverse engineering the same BDD-based reversible circuits used in the previous experiment, however, giving that the attacker knows the applied synthesis approach.

We conclude that the attacker can identify the majority of the constant input bits value of BDD-based reversible circuits. On average $81.6\%$ of the ancillary input bits can be recovered by the attacker. Thus, the attacker can identify most of the target function. The large number of possible embeddings in column 9 is due to the large number of potential primary outputs. Although the attacker can recover the location of primary outputs that satisfy the second property of BDD-based reversible circuits in Section~\ref{Attack Procedure on BDD-based Reversible Circuits}, he/she can not determine whether the outputs of sub-circuits, which are connected to the reversible circuit outputs and also used to control other target lines, are garbage outputs. Thus, the number of all possible primary output combinations increases significantly depending on the size of these potential output bits. {\bf Our experiment shows that the attacker can identify all the primary outputs of the given BDD-based reversible circuits using our proposed reverse engineering except for three reversible circuits}. In other words, the number of possible embeddings of the BDD-based reversible circuits can be reduced significantly by ignoring the potential primary outputs, which makes reverse engineering easier.

\begin{figure}[htbp]
\centering
\subfigure []{\label{fig:com_e_}\includegraphics[width=3.5in]{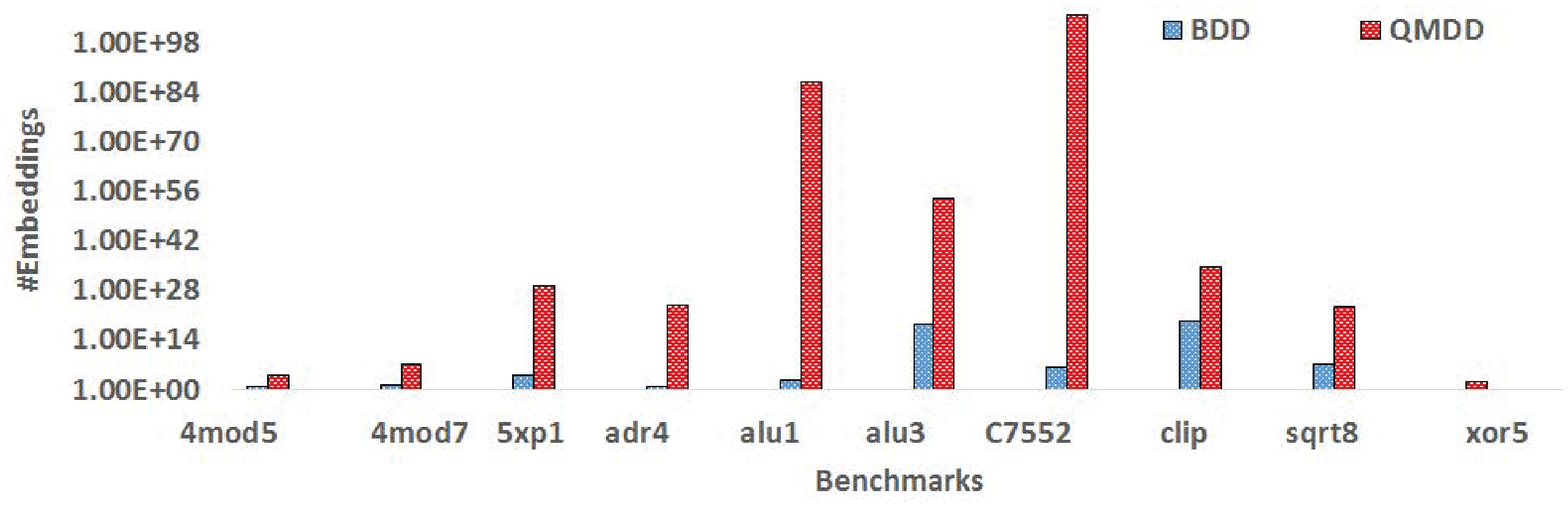}}\\ 
\subfigure []{\label{fig:com_q}\includegraphics[width=3.5in]{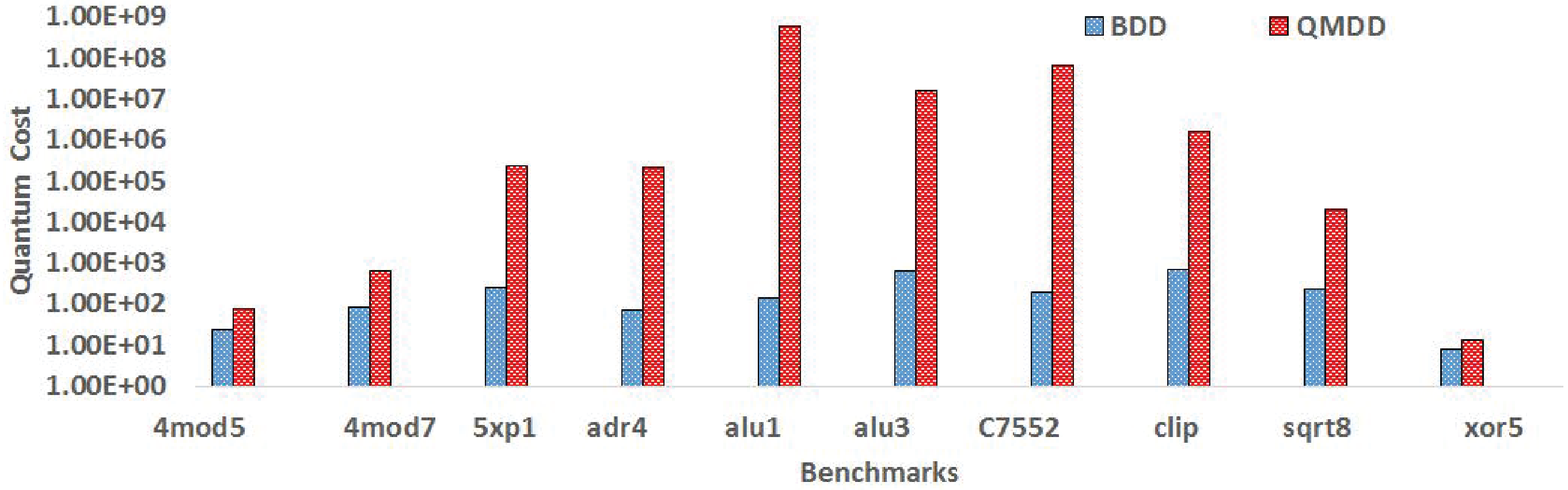}}               
\caption{Comparison between QMDD- and BDD-based reversible circuit in terms of (a) the number of embedding and (b) the quantum cost. }
\label{fig:comp_e_q}
\end{figure}

Fig.~\ref{fig:comp_e_q} shows a comparison between BDD- and QMDD-based (with 0-embedding) reversible circuits in terms of the number of embeddings and the quantum cost of a selected set of target functions given that the attacker knows the applied synthesis approach. Reverse engineering BDD-based reversible circuits generates less number of embeddings, and thus, it is easier to attack, compared to QMDD-based reversible circuits. On the other hand, the quantum cost of the QMDD-based reversible circuits exceeds the corresponding one of BDD-based reversible circuits.

\subsection{Reverse Engineering Reversible Circuits with Input/Output Scrambling}
\begin{figure*}[t]
\centering
\subfigure []{\label{fig:obs_BDD_Q}\includegraphics[width=3.5in]{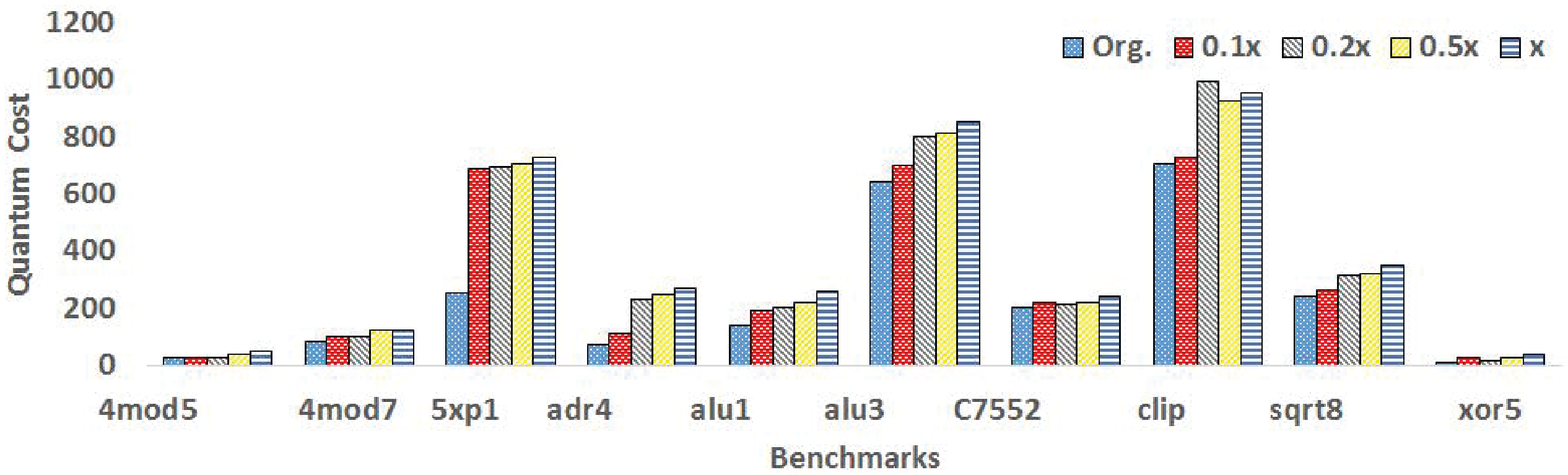}}\hspace{0.5em}
\subfigure []{\label{fig:obs_BDD_EMD1}\includegraphics[width=3.5in]{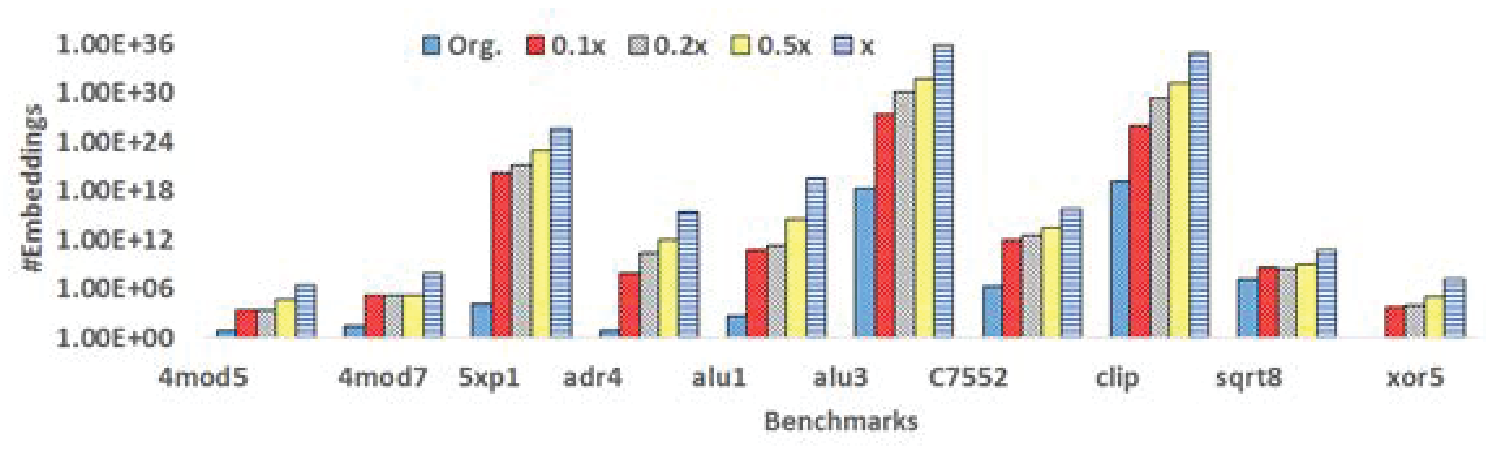}}               
\subfigure []{\label{fig:obs_QMDD_Q}\includegraphics[width=3.5in]{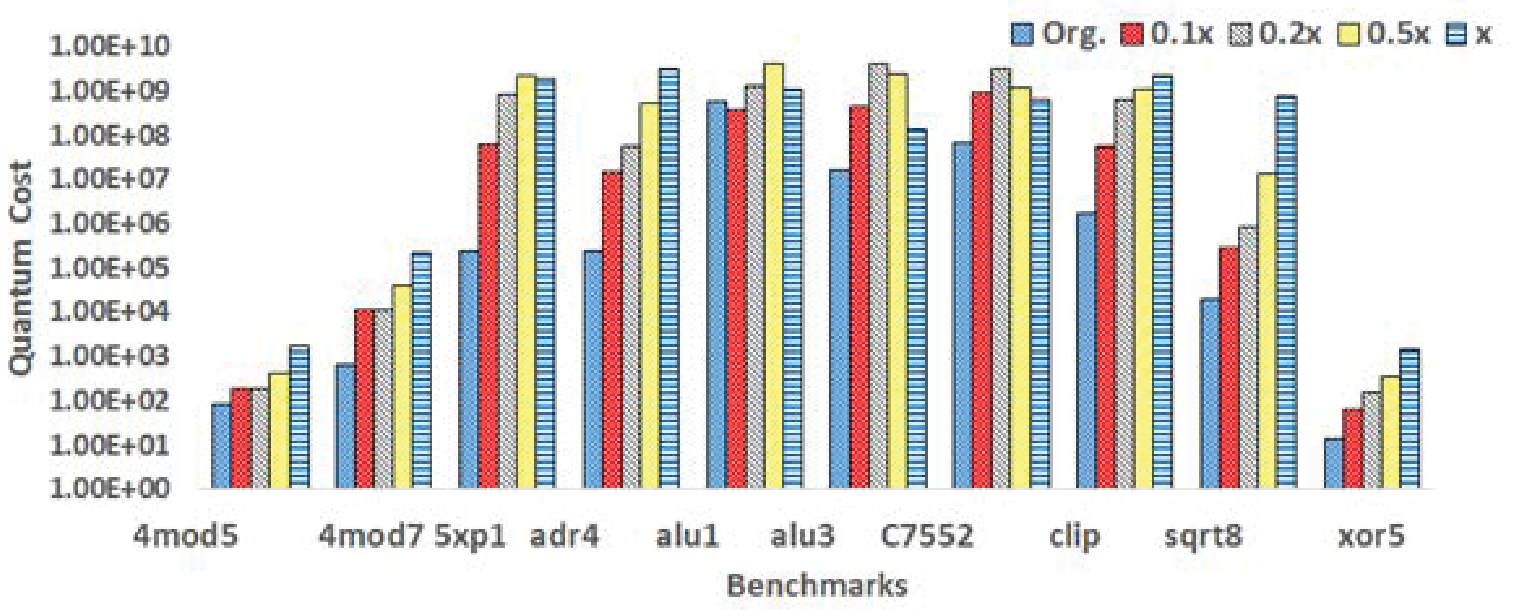}}\hspace{0.5em}
\subfigure []{\label{fig:obs_QMDD_EMD}\includegraphics[width=3.5in]{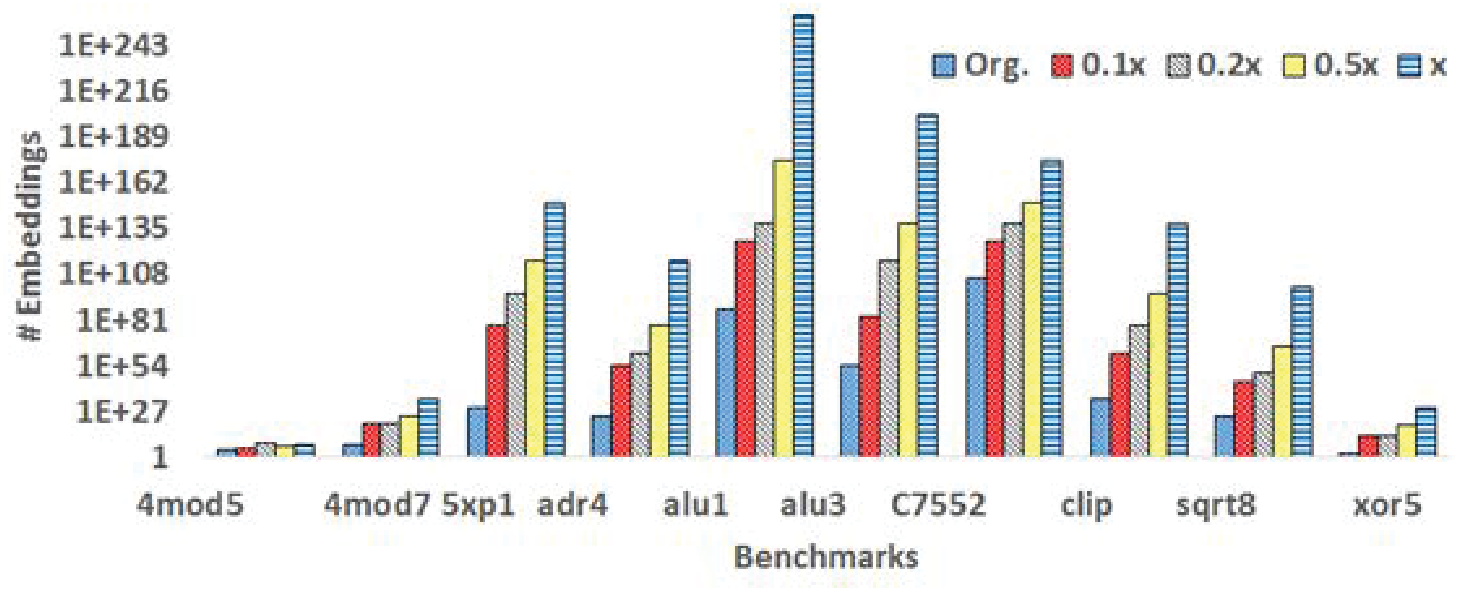}}   
\caption{Obfuscation using extra ancillary inputs. In BDD-based reversible circuits (a) Quantum Cost and b) number of embeddings. In QMDD-based reversible circuits (c) Quantum Cost and b) number of embeddings.}
\label{fig:obf_QMDD_BDD}
\end{figure*}

We apply our attacks on BDD- and QMDD-based reversible circuits in the presence of extra ancillary inputs and garbage outputs. In Fig.~\ref{fig:obf_QMDD_BDD}, for each $n$ input/output reversible circuit, we created five variants of the reversible circuit with 0, $0.1x$, $0.2x$, $0.5x$, and $x$ extra ancillary inputs, where $x$ is the number of input/output bits of the reversible circuit\footnote{The number of extra ancillary input bits varies from 1 to 21.}.

Additional ancillary inputs significantly inflate the number of embeddings of not only QMDD-based reversible circuits but also BDD-based reversible. Every input classified as a primary input by reverse engineering BDD-based reversible circuit can be a potential ancillary input in the presence of our proposed input scrambling approach, which results in a massive increase in the number of embeddings. Moreover, increasing the number of the extra ancillary inputs in few BDD-based reversible circuits may result in a slight reduction in the quantum cost and the number of embeddings due to the random value of the ancillary input bits that determines the corresponding sub-circuits. Recall that Shannon decomposition used in BDD-based reversible circuits generates different sub-circuits, which have different numbers of gates. Some of these sub-circuits can easily be used to infer the associated ancillary input bit value, while others are difficult.

\begin{figure*}[!tb]
\centering
\subfigure []{\label{fig:obsg_BDD_Q}\includegraphics[width=3.5in]{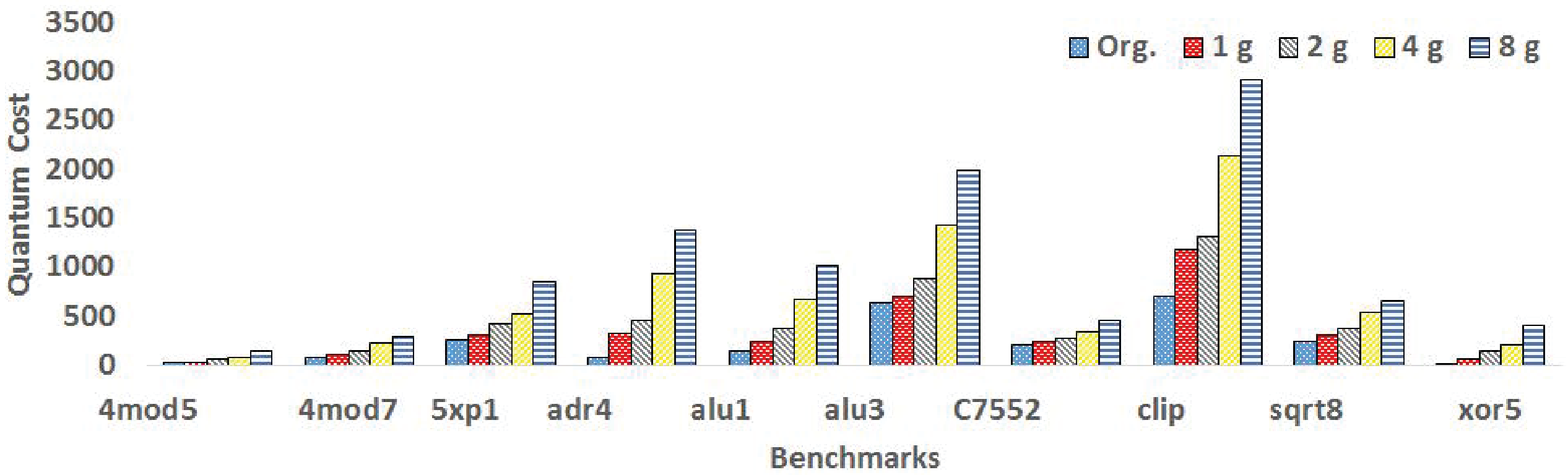}}\hspace{0.5em}
\subfigure []{\label{fig:obsg_BDD_EMD1}\includegraphics[width=3.5in]{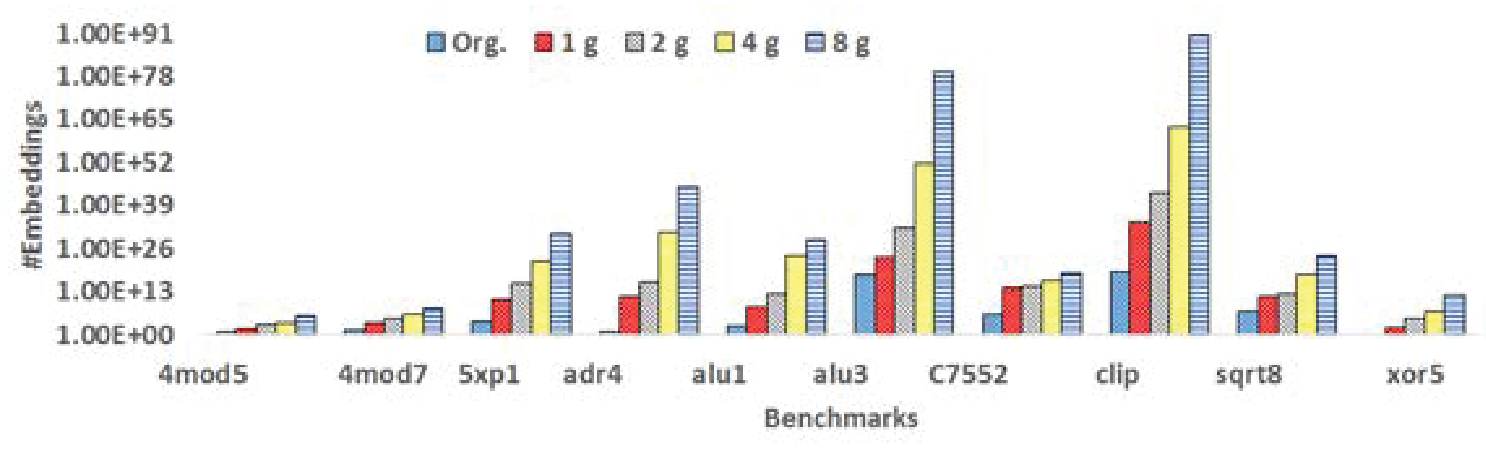}}               
\subfigure []{\label{fig:obsg_QMDD_Q}\includegraphics[width=3.5in]{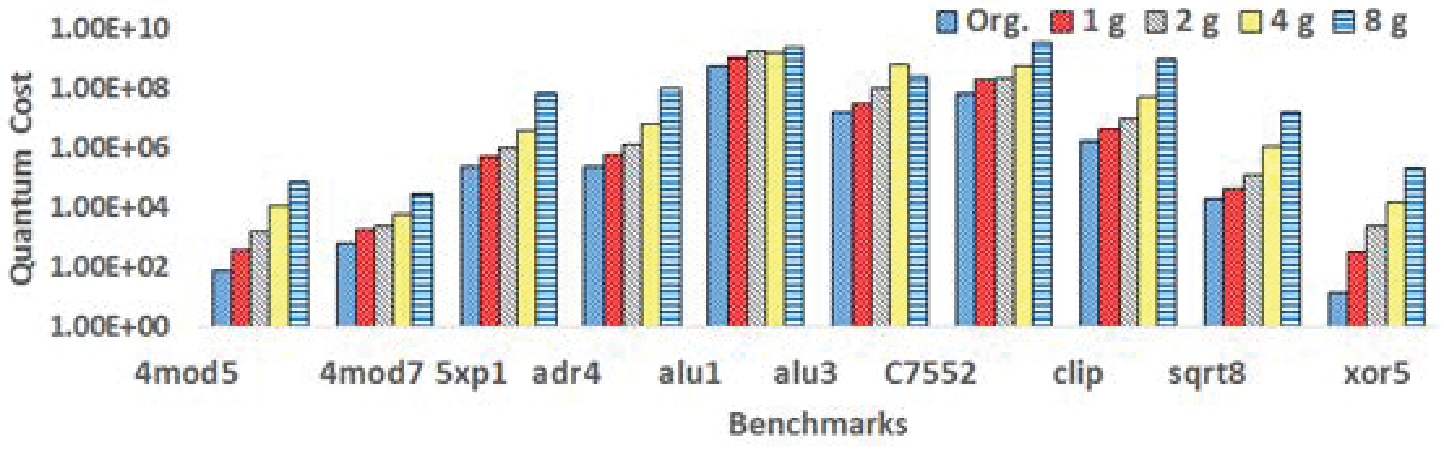}}\hspace{0.5em}
\subfigure []{\label{fig:obsg_QMDD_EMD}\includegraphics[width=3.5in]{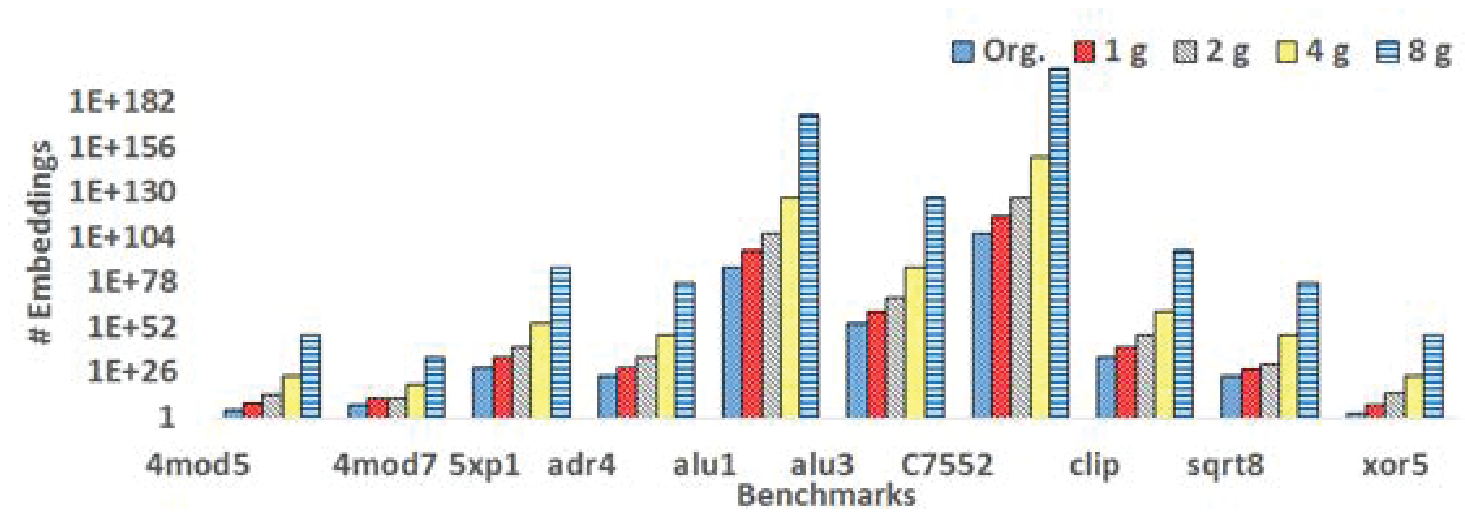}}   
\caption{Obfuscation using extra garbage outputs. In BDD-based reversible circuits (a) Quantum Cost and (b) number of embeddings. In QMDD-based reversible circuits  (c) Quantum Cost and (b) number of embeddings.}
\label{fig:obfg_QMDD_BDD}
\end{figure*}

In the presence of additional garbage outputs in a BDD-based reversible circuit, the first step of the attack can not be applied anymore. As shown in Fig.~\ref{fig:obfg_QMDD_BDD}, output scrambling of BDD-based reversible circuits results in a larger number of possible embeddings compared to input scrambling at the expense of high quantum cost. 

While we show the impact of our proposed input/output scrambling on QMDD-based reversible circuits in Fig.~\ref{fig:obs_QMDD_EMD},~\ref{fig:obs_QMDD_Q},~\ref{fig:obsg_QMDD_EMD}, and~\ref{fig:obsg_QMDD_Q}, we emphasize that reverse engineering QMDD-based reversible circuits without applying our proposed input/output scrambling is still difficult.

\section{Conclusion}
In this paper, we analyze reverse engineering of reversible logic. We focus on non-reversible functions embedded into reversible circuits. We measure the number of possible embeddings of reversible circuits implemented using different synthesis approaches and under different threat models as a security metric to evaluate the difficulty of extracting the target function through reverse engineering. We propose an attack to reverse engineering reversible circuits generated using BDD-based synthesis approach, an example of structural synthesises, that provides scalability at the cost of significant circuit lines overhead. We show that reversible circuits created using functional synthesis approaches are inherently secure against reverse engineering at the expense of high quantum cost. We also propose a countermeasure to thwart reverse engineering of BDD-based reversible circuits by adding extra ancillary inputs or garbage outputs prior to the synthesis step to scramble the inputs or the outputs of reversible circuits, respectively. 

Our future work will explore reverse engineering of reversible circuits generated using synthesis approaches that provide both scalability and significant reduction in the number of circuit lines. Another direction is to develop a systematic way to identify the synthesis approach that generates a given reversible circuit considering all the state-of-the-art synthesis approaches.   




%
\bibliographystyle{IEEEtran}
\bibliography{mybib}

\end{document}